\documentclass[iop]{emulateapj}

\slugcomment{Accepted for publication in The Astrophysical Journal.}
\shortauthors{Homan et al.}
\shorttitle{MOJAVE. XII. Acceleration of Blazar Jets}
\begin{document}
\title{MOJAVE XII: Acceleration and Collimation of Blazar Jets on Parsec Scales}
\author{
D. C. Homan\altaffilmark{1},
M. L. Lister\altaffilmark{2},
Y. Y. Kovalev\altaffilmark{3,4},
A. B. Pushkarev\altaffilmark{5,6,4},
T. Savolainen\altaffilmark{4,7}
K. I. Kellermann\altaffilmark{8},
J. L. Richards\altaffilmark{2},
and
E. Ros\altaffilmark{4,9,10}
}

\altaffiltext{1}{
Department of Physics, Denison University, Granville, OH 43023;
\email{homand@denison.edu}
}
\altaffiltext{2}{
Department of Physics, Purdue University, 525 Northwestern Avenue,
West Lafayette, IN 47907, USA;
}

\altaffiltext{3}{
Astro Space Center of Lebedev Physical Institute,
Profsoyuznaya 84/32, 117997 Moscow, Russia;}
\altaffiltext{4}{
Max-Planck-Institut f\"ur Radioastronomie, Auf dem H\"ugel 69,
53121 Bonn, Germany;
}
\altaffiltext{5}{
Pulkovo Observatory, Pulkovskoe Chaussee 65/1, 196140 St.
Petersburg, Russia;
}
\altaffiltext{6}{
Crimean Astrophysical Observatory, 98409 Nauchny, Crimea, Ukraine;}
\altaffiltext{7}{
Aalto University Mets\"ahovi Radio Observatory, Mets\"ahovintie 114, 02540 Kylm\"al\"a, Finland;
}
\altaffiltext{8}{
National Radio Astronomy Observatory, 520 Edgemont Road, Charlottesville, VA 22903, USA;
}
\altaffiltext{9}{
Observatori Astron\`omic, Universitat de Val\`encia,
  Parc Cient\'{\i}fic, C. Catedr\'atico Jos\'e Beltr\'an 2, E-46980
  Paterna, Val\`encia, Spain}

\altaffiltext{10}{
Departament d'Astronomia i Astrof\'{\i}sica,
  Universitat de Val\`encia, C. Dr. Moliner 50, E-46100 Burjassot,
  Val\`encia, Spain}

\begin{abstract}
We report on the acceleration properties of 329 features in 95
blazar jets from the MOJAVE VLBA program. Nearly half the features 
and three-quarters of the jets show significant changes in speed 
and/or direction. In general, apparent speed changes are 
distinctly larger than changes in direction, indicating that changes 
in the Lorentz factors of jet features dominate the observed speed 
changes rather than bends along the line of sight. 
Observed accelerations tend to increase the speed of features 
near the jet base, $\lesssim 10-20$ parsecs projected,
and decrease their speed at longer distances.  The range of apparent 
speeds at fixed distance in an individual jet can span a factor of a few, 
indicating that shock properties and geometry may influence the 
apparent motions; however, we suggest that the broad trend of jet
features increasing their speed near the origin is due to an 
overall acceleration of the jet flow out to de-projected 
distances of order $10^2$ parsecs, beyond which the flow begins to 
decelerate or remains nearly constant in speed.  We estimate 
intrinsic rates of change of the Lorentz factors in the galaxy
frame of order $\dot{\Gamma}/\Gamma \simeq 10^{-3}$ to $10^{-2}$ 
per year which can lead to total Lorentz factor changes of a 
factor of a few on the length scales observed here.
Finally, we also find evidence for jet collimation at 
projected distances of $\lesssim 10$ parsecs in the form of
the non-radial motion and bending accelerations
that tend to better align features with the inner jet.
\end{abstract}

 \ 

\section{Introduction}

The National Radio Astronomy Observatory's Very Long Baseline Array (VLBA)\footnote{
The National Radio Astronomy Observatory is a facility of the National Science 
Foundation operated under cooperative agreement by Associated Universities, Inc.} 
has revolutionized the study of powerful extragalactic radio jets associated
with Active Galactic Nuclei (AGNs) by allowing highly sensitive, regular observations of
the parsec-scale structure and polarization of large samples. The extraordinary resolution
of the VLBA combined with the extreme time compression created by 
highly relativistic motion beamed along our line of sight
allows study on a human timescale of physical processes spanning centuries 
in the host galaxy: $\Delta t_\mathrm{obs} = (1+z)(1-\beta\cos\theta)\Delta t$, where
$\Delta t_\mathrm{obs}$ is the observed time interval, $\Delta t$ is the time in the host galaxy,
$\beta$ is the speed of the moving jet feature in units of the speed of light, and $\theta$
is the angle the jet feature makes to our line of sight.   This ability to directly observe
 long timescale processes over the span of a decade is essential for addressing long standing 
questions about the formation, acceleration, and collimation of AGN jets from the supermassive
black hole/accretion disk system \citep[e.g.][]{Meier01,VK04,S05,K11}. 

In recent years, Very Long Baseline Interferometry (VLBI) monitoring programs of the 
parsec-scale structure of 
powerful extragalactic radio jets have begun to span long enough {\em time} baselines
to directly study the apparent acceleration of large numbers of individual jet features 
across many jets \citep{LCH09,H09,P12}.  These works 
extend earlier studies of acceleration in individual jets \citep[e.g.][]{WCRB94,Z95,G01,H03,JML04,SWVT06} 
or small numbers of 
jets \citep{H01,J05,PMF07} to allow statistical analysis of the patterns in observed acceleration. 
The MOJAVE program\footnote{http://www.astro.purdue.edu/MOJAVE/} \citep[e.g.][]{LH05,LAA09}
is a continuation of the 2cm Survey \citep{K98,KL04,Z02,K05}, and together they form 
the longest running VLBI astrophysical monitoring program with continuous observations of many jets. 
To date, eleven survey papers have been published in the 
MOJAVE series. Our most recent kinematics results were presented by \citet[][hereafter Paper X]{LAA13}
for 887 moving features in 200 AGN jets based on observations spanning the period from 
1994 Aug 31 to 2011 May 1.  Paper X updated our earlier kinematic and acceleration results 
presented in \citet[][hereafter Paper VI]{LCH09} and \citet[][hereafter Paper VII]{H09}, extended 
our time baseline by four years, increased our sample of jet features by nearly 70\%, and examined
changes in jet position angle over time and the dispersion in jet feature speed within individual jets. 

In this paper, we analyze observed accelerations for the kinematic results presented in Paper X.
This work extends the analysis developed in Paper VII by expanding our sample of high
quality motions suitable for detailed acceleration study from 203 features in 63 jets to 329 
features in 95 jets.  We study both {\em parallel} accelerations, along the direction of motion,
and {\em perpendicular} accelerations that change the direction of motion; however, we focus particular 
attention on the relationship between parallel acceleration and distance from the base of the jet.  
Paper VII reported 
that jet features tend to accelerate at short projected distances but decelerate at large distances, 
suggesting that the jet flow was speeding up near the base and decelerating further out, with a
transition region in the range $\sim 15$ pc.  In their analysis of ten years of observations performed by 
the global VLBI array for astrometry and geodesy \citep{P09}, \citet{P12} confirmed the preponderance of positive 
accelerations found at short distances but could not confirm the switch to a preponderance of 
decelerations at larger distances from the core since their sample had too few jet features beyond 15 pc.
In addition to revisiting this relationship with much improved data, we also
examine directly the relationship between apparent speed and jet distance through the use of
{\em speed profiles} predicted from the acceleration kinematics fits, and we discuss the
implications for changes in the jet Lorentz factor on parsec scales.   

This paper is organized as follows:  Section 2 reviews the relationship between observed
accelerations and intrinsic changes in speed and direction of parsec scale jet features. 
The sample of jet features studied for
acceleration is described in section 3, where we also report the results of our analysis
of these data.  Section 4 presents a discussion of these results, and our conclusions appear in \S{5}. 
Throughout this paper we assume a cosmology with $H_0=71$ km s$^{-1}$ Mpc$^{-1}$, 
$\Omega_\Lambda = 0.73$, and $\Omega_M=0.27$ \citep{2009ApJS..180..225H}.

\section{Accelerated Motion} 

Here we summarize relationships describing changes in apparent speed of parsec-scale AGN 
jet features and how those apparent changes relate to changes in intrinsic properties of
the jet. More detailed discussion and derivations of these expressions are given in Paper VII.

The apparent speed on the 2-D sky plane for a jet feature or pattern moving with intrinsic speed $\beta$ at
angle $\theta$ to the line of sight is given by the following familiar expression:

\begin{equation}
\label{e:beta_app}
\beta_\mathrm{app} = \frac{\beta\sin\theta}{1-\beta\cos\theta}.
\end{equation}

The intrinsic velocity, $\vec{\beta}$, is a vector which can change in speed and/or direction, resulting 
in apparent changes to the motion in the 2-D sky plane.  We characterize these apparent changes as
accelerations either {\em parallel} or {\em perpendicular} to the apparent velocity:

\begin{equation}
\label{e:dbeta_app1}
\dot{\beta}_{\parallel app} = \frac{d\beta_{\parallel app}}{dt_\mathrm{app}} = \frac{\dot{\beta}\sin\theta+\beta\dot{\theta}(\cos\theta-\beta)}{(1-\beta\cos\theta)^3}
\end{equation}

\noindent and

\begin{equation}
\label{e:dbeta_app2}
\dot{\beta}_{\perp app} = \frac{d\beta_{\perp app}}{dt_\mathrm{app}} =
\frac{\beta\dot{\phi}\sin\theta}{(1-\beta\cos\theta)^2},
\end{equation}

\noindent where $\dot{\beta}$, $\dot{\theta}$, and $\dot{\phi}$ are the intrinsic rates of change of the jet feature's speed, 
angle to the line of sight, and azimuthal angle of the component's motion respectively.  
Note that the intrinsic velocity 
vector angles, $\theta$ and $\phi$, are as defined in Figure A1 of Paper VII, and that $\phi$ corresponds directly to 
observed proper motion vector direction for jet features given in \S{3}.  

From these expressions, it is clear that parallel accelerations, i.e. speeding up or slowing down along the apparent
velocity vector, can be generated {\em either} by changes in the intrinsic speed, $\beta$, {\em or} by changes in the
angle to the line of sight, $\theta$, or both.  Determining the extent to which intrinsic speed changes
can explain the observed accelerations is crucial to their physical interpretation.  If all of the observed
parallel accelerations are due only to intrinsic changes in speed, at any given time 
the relative rate of change of the Lorentz factor,
$\Gamma = 1/\sqrt{1-\beta^2}$, is given by the following expression (Paper VII):

\begin{equation}
\frac{\dot{\Gamma}}{\Gamma} = \frac{\beta^2}{\delta^2}\frac{\dot{\beta}_{\parallel app}}{\beta_\mathrm{app}}
\end{equation}

\noindent where $\delta$ is the Doppler factor: $\delta = 1/(\Gamma(1-\beta\cos\theta))$.

However, if all the intrinsic changes are changes in direction only, i.e. ``bending'' of
the component trajectory, then apparent parallel accelerations should usually be accompanied 
by apparent perpendicular accelerations.  Indeed,
we can use the relative magnitudes of the parallel and perpendicular accelerations to draw
statistical conclusions about the prevalence of intrinsic changes in speed in our sample.  Paper VII 
showed that for a parsec-scale flux-density limited sample, like MOJAVE, the parallel accelerations
should only be about $60$\% of the magnitude of the perpendicular accelerations when
averaged across the sample if the observed accelerations are due only to intrinsic 
changes in direction of motion.  

\section{Data and Results}
\label{s:results}

Paper X presented parsec-scale kinematics results for 887 moving features
in 200 AGN jets, including acceleration fits for 557 features with 10 or more epochs of 
observation. Here we analyze the apparent accelerations of jet features both 
parallel, $\dot{\mu}_\parallel$, and perpendicular, $\dot{\mu}_\perp$, to their 
observed proper motion vector, $\vec{\mu}$.  As in our original acceleration analysis in 
Paper VII, we restrict the sample for this paper to those having a proper 
motion, $\mu$, of at least $3\sigma$ significance,
a known redshift (required for relative acceleration analysis, see below), and an
uncertainty of no more than $5^\circ$ in the misalignment between the average 
position angle of the feature over time, $\langle\vartheta\rangle$, and the direction
of its proper motion vector, $\phi$.  This last condition guarantees that the meaning of ``parallel''
and ``perpendicular'', defined relative to the observed proper motion vector,
is unambiguous. Applying these criteria yields a sample of 329 features in 95 jets  
suitable for the analysis in this paper, and they are listed in Table 1, along
with a summary of their kinematic properties.
Paper X includes 
the full proper motion results, along with plots of their motion, for all of these 
features.\footnote{Paper X does not report 
an apparent speed or relative accelerations for 0716$+$714 due to a lack of
a spectroscopic redshift.  Here we compute results for 0716$+$714 using the 
redshift estimate, $z=0.31\pm0.08$ by \citet{NPS08} based on host galaxy magnitude. 
Using a different value in the range, $z=0.2315-0.322$ given by \citet{DNF13} would not 
change our acceleration results appreciably.}
It is important to note that we can only confidently measure and compare the
acceleration properties of the well defined, long lived features meeting the criteria
described above and in Paper X.  Jets also exhibit complex behavior, including stationary
and transitory features, that may not be captured by the features suitable for acceleration study.  
We include discussion of the complexities in deducing the jet flow behavior from observed 
features in \S{\ref{s:shocks}}.

To compare accelerations of jet features with different apparent speeds, we 
use their relative accelerations as defined in Paper VII:

\begin{equation}
\label{e:relative_accel}
\dot{\eta}_\parallel = \dot{\beta}_{\parallel app}/\beta_\mathrm{app} = (1+z)\dot{\mu}_\parallel/\mu ,
\end{equation}
and
\begin{equation}
\dot{\eta}_\perp = \dot{\beta}_{\perp app}/\beta_\mathrm{app} = (1+z)\dot{\mu}_\perp/\mu .
\end{equation}

\begin{figure}
\includegraphics[scale=0.45,angle=0,totalheight=0.45\textheight]{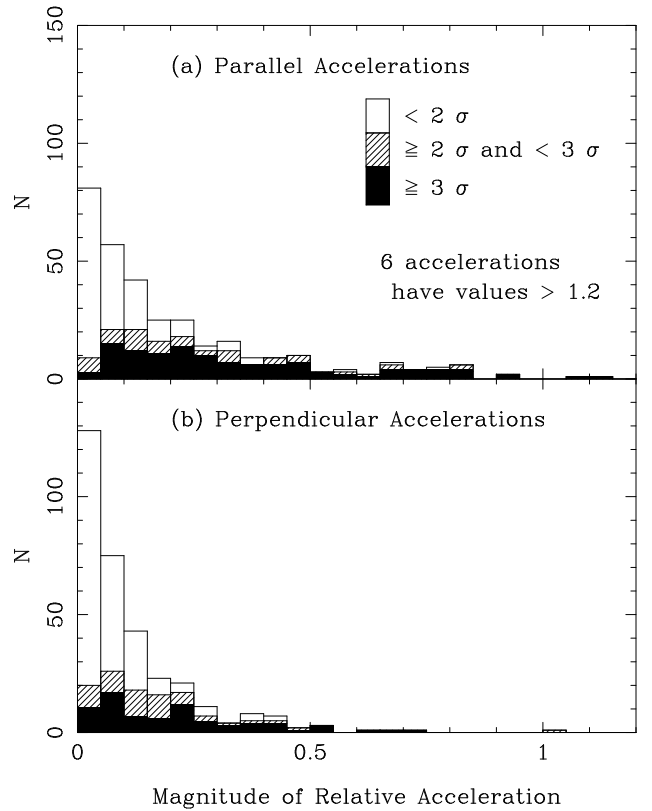}
\figcaption{\label{f:accel_hist}
Histograms of magnitudes of relative accelerations, parallel, $\dot{\eta}_\parallel$ (panel (a)), 
and perpendicular, $\dot{\eta}_\perp$ (panel (b)), to the proper motion vector direction.  Hash and solid
fill styles indicate angular acceleration significant at the $2-3\sigma$ and $\geq 3\sigma$ levels,
respectively.  As indicated, six parallel accelerations lie to 
the right of the boundary of panel (a).
}
\end{figure}

Table 2 summarizes the properties of the acceleration sample as a whole, and 
the distributions of the parallel and perpendicular relative accelerations are
plotted in Figure \ref{f:accel_hist}.  More than one-third ($37$\%) of the jet
features in our sample have significant ($\geq 3\sigma$) parallel accelerations, 
indicating a change in apparent speed, and nearly a quarter ($23$\%) have
significant perpendicular accelerations, indicating a change in direction of
apparent motion. Altogether, one-half ($50$\%) of the jet features have one
or both kinds of acceleration at the $\geq 3\sigma$ level, and three-quarters
of the 95 jets we studied have at least one feature with significant acceleration.  
In general, parallel accelerations are distinctly larger
than perpendicular accelerations, and a Kolmogorov-Smirnov (K$-$S) test gives
a probability $P < 10^{-7}$ $(N=329)$ that they are drawn from the same distribution. 
The typical magnitude of the observed parallel accelerations is nearly a 
factor of two larger than the perpendicular accelerations, irrespective
of whether the means or the medians are used.

\begin{deluxetable}{lc}
\tablecolumns{1}
\tablewidth{0pc}
\tabletypesize{\scriptsize}
\tablenum{2}
\tablecaption{Statistics of Acceleration Sample\label{t:accel_sample}}
\tablehead{\colhead{Property} & \colhead{Value}}
\startdata
Number Jet Features                   & $329$\\
Number of $\geq 3\sigma$ Parallel Accelerations       & $123$\\
Number of $\geq 3\sigma$ Perpendicular Accelerations  & $76$\\
Mean Magnitude of Relative Parallel Accel.            & $0.233$ yr$^{-1}$ \\
Median Magnitude of Relative Parallel Accel.          & $0.130$ yr$^{-1}$ \\
Mean Magnitude of Relative Perp. Accel.            & $0.115$ yr$^{-1}$ \\
Median Magnitude of Relative Perp. Accel.          & $0.071$ yr$^{-1}$ \\
Number of High Parallel Accel. ($|\dot{\eta}_\parallel| - 2\sigma > 0.1$) & $85$ \\
Number of Low Parallel Accel. ($|\dot{\eta}_\parallel| + 2\sigma \leq 0.1$) & $54$ \\
Number of High Perp. Accel. ($|\dot{\eta}_\perp| - 2\sigma > 0.1$) & $34$ \\
Number of Low Perp. Accel. ($|\dot{\eta}_\perp| + 2\sigma \leq 0.1$) & $83$ \\
Number of $\geq 3\sigma$ Non-radial Motions       & $156$ \\
Number of Non-radial Motions with $|\langle\vartheta\rangle-\phi| \geq 10^\circ$  & $87$ 
\enddata
\end{deluxetable}

\begin{figure}
\includegraphics[scale=0.4,angle=-90,totalheight=0.265\textheight]{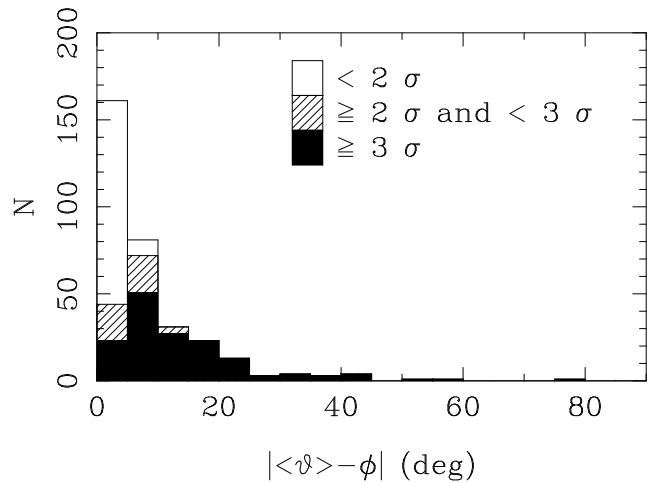}
\figcaption{\label{f:nonradial_motion_hist}
Histogram of the magnitude of the proper motion misalignment angle, $|\langle\vartheta\rangle-\phi|$.
Hash and solid fill styles indicate non-radial motions significant at the $2-3\sigma$ and $\geq 3\sigma$ levels
respectively. Three jet features had misalignment angles $\geq 90^\circ$ and do not appear in the plot.
}
\end{figure}

In addition to the directly observed accelerations, we measure the non-radial
motion of jet features, comparing their mean structural position angle, $\langle\vartheta\rangle$,
to their proper motion vector direction, $\phi$. Non-radial features have a 
proper motion vector that does not point back to the jet core (see the next
subsection).
Figure \ref{f:nonradial_motion_hist}
is a histogram of these observed mis-alignments.  Nearly half of our sample
($47$\%) has motion that is significantly non-radial, and one-quarter ($26$\%)
of the motions are mis-aligned by $10^\circ$ or more. 

\subsection{Parallel Acceleration vs. Jet Distance}

As we explore in \S{\ref{s:discuss}}, the distinctly larger magnitudes of
parallel accelerations as compared to perpendicular accelerations
indicate that the observed parallel accelerations are more likely 
to reflect intrinsic changes in the Lorentz factor of propagating features than 
changes in their angle to the line of sight.  In this section we 
explore the relationship between significant ($\geq 3\sigma$) 
parallel accelerations and distance of the feature from the 
observed base of the jet or ``core''.\footnote{We note that 
NGC 1052 (0238-084) at a redshift of just $0.005$ is significantly 
closer than the other jets in this sample and has projected linear
distances an order of magnitude smaller. We exclude NGC 1052 
jet features at $< 1$ parsec from the plots in this section to 
allow a clearer view of the bulk of the sample, but they are included
in our statistics.}     

\begin{figure*}
\begin{center}
\includegraphics[scale=0.45,angle=0,totalheight=0.72\textheight]{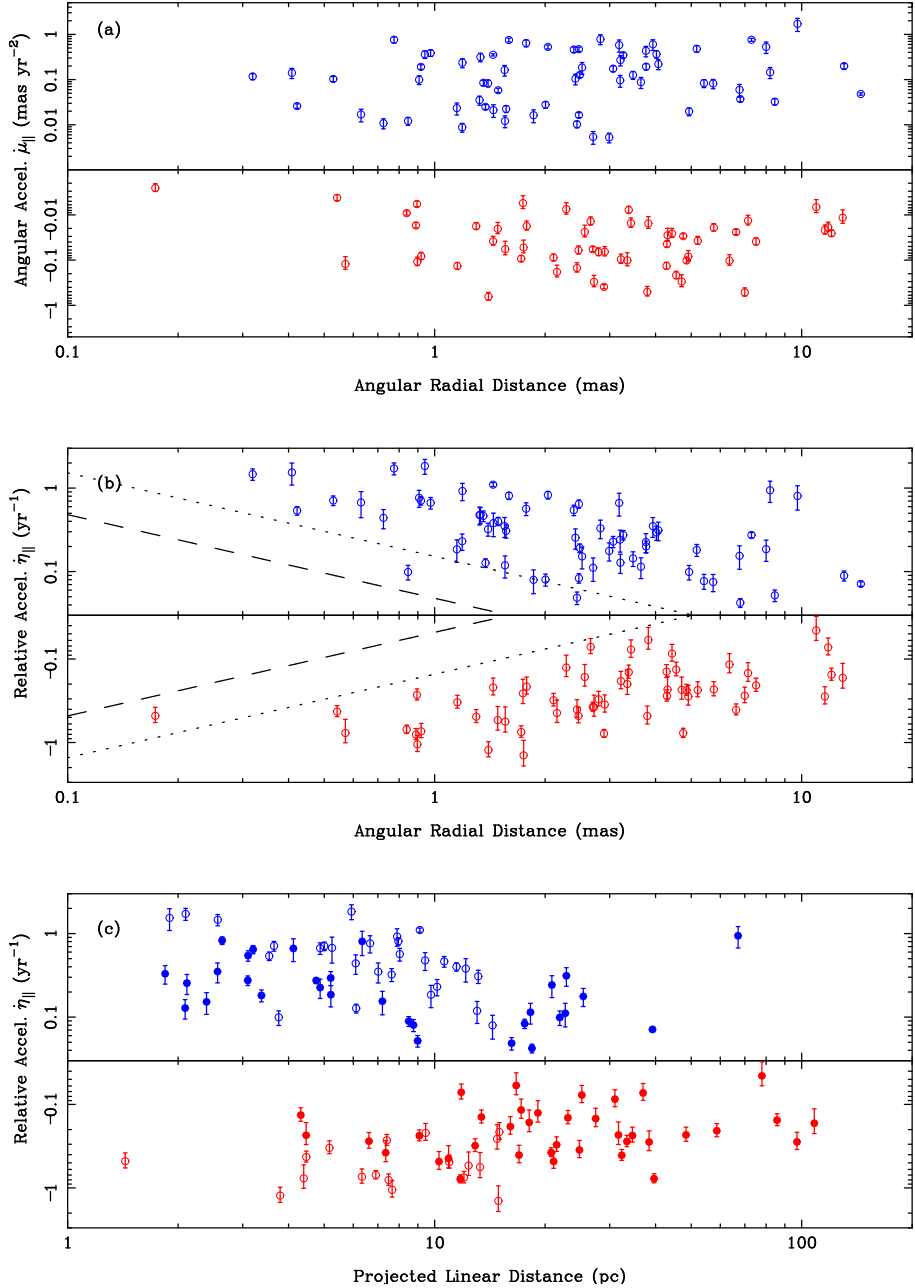}
\end{center}
\figcaption{\label{f:R_dist_accel_plot}
Panels (a) and (b) plot significant parallel accelerations ($\geq 3\sigma$) against average 
angular distance from the core position in milli-arcseconds.  Panel (a) plots measured angular 
acceleration, $\dot{\mu}_\parallel$,
and panel (b) plots relative parallel acceleration, $\dot{\eta}_\parallel$. The dashed and dotted lines
in panel (b) indicate estimates for our threshold for measuring large {\em} relative accelerations
at small angular distance, see \S{3.1}.  Panel (c) plots relative parallel acceleration
versus projected linear distance in parsecs.  The solid points indicate features at angular
distances $\geq 2.0$ milli-arcseconds.  Note that in
panel (c), eight features from NGC 1052 at projected linear distances $< 1.0$ pc are not
plotted for clarity; seven of these have positive acceleration, and one has negative.
}
\end{figure*}

Figure \ref{f:R_dist_accel_plot} plots the magnitude of positive (blue) and
negative (red) parallel accelerations as a function of angular distance
along the jet in panels (a) and (b).  Panel (a) plots the measured
{\em angular} accelerations, $\dot{\mu}_\parallel$, directly, and
panel (b) plots the {\em relative} accelerations, $\dot{\eta}_\parallel$,
as defined in equation \ref{e:relative_accel}. In panel (b), there
appears to be a trend of decreasing relative accelerations with increasing
angular separation up to about $\langle R\rangle \simeq 2$ milli-arcsec; however, this is
an artifact of calculating relative acceleration by dividing out the
average angular speed, $\mu$.  Jet features at small average angular 
separation must have a small $\mu \leq 2\langle R\rangle/\Delta t_{obs}$,
where $\Delta t_{obs}$ is the time span of our observations. 
To be measured at the three sigma level, they must have a relative 
acceleration at least $\dot{\eta}_\parallel \gtrsim 3\sigma\Delta t_{obs}/2\langle R\rangle$
where $\sigma$ is the uncertainty in the measured angular acceleration.  
Median values for
the quantities are $\sigma=0.0135$ mas yr$^{-2}$ and $\Delta t_{obs} = 7.5$ years, and
the dotted line in panel (b) shows our approximate measurement threshold
using these median values.  Individual measurements may have smaller
values of these quantities separately, but their product, $\sigma\Delta t_{obs}$, 
is $> 0.032$ mas yr$^{-1}$ in 90\% of the jet features we studied, and the
dashed line in panel (b) gives a more conservative threshold using this value.

\begin{figure}
\includegraphics[scale=0.45,angle=0,totalheight=0.49\textheight]{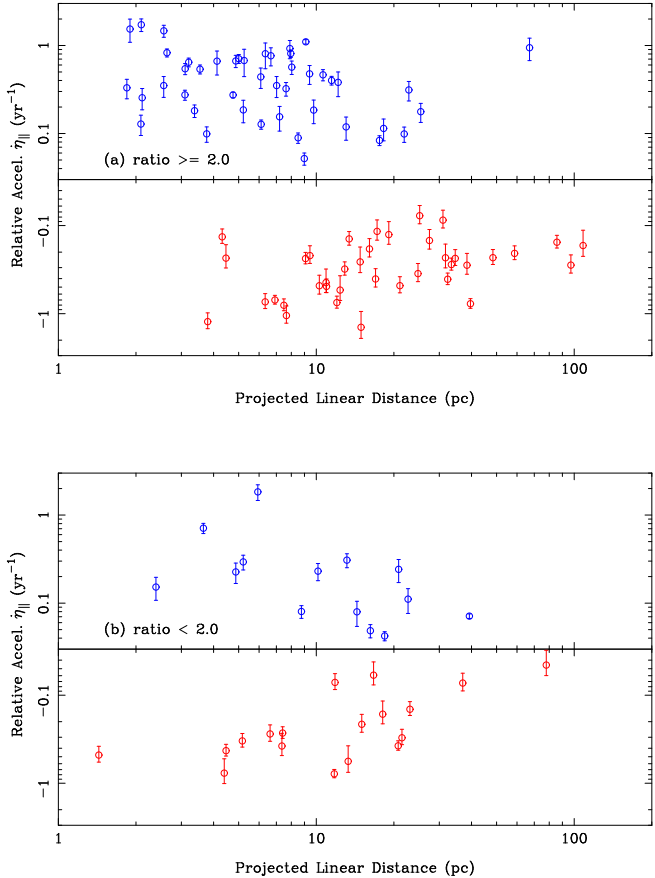}
\figcaption{\label{f:accel_v_pc_ratio}
Relative parallel accelerations, $\dot{\eta}_\parallel$ for all jet features with $\geq 3\sigma$ 
parallel accelerations plotted against average projected linear distance from the core 
position in parsecs.  Features with parallel to perpendicular acceleration ratios of $\geq 2.0$ 
are plotted in panel (a), and features with ratios $< 2.0$ are plotted in panel (b).
Eight features from NGC 1052 at projected distances $< 1.0$ pc do not appear in the plots 
for clarity; seven of those have positive acceleration, and one has negative.
}
\end{figure}

\begin{figure}
\includegraphics[scale=0.45,angle=0,totalheight=0.49\textheight]{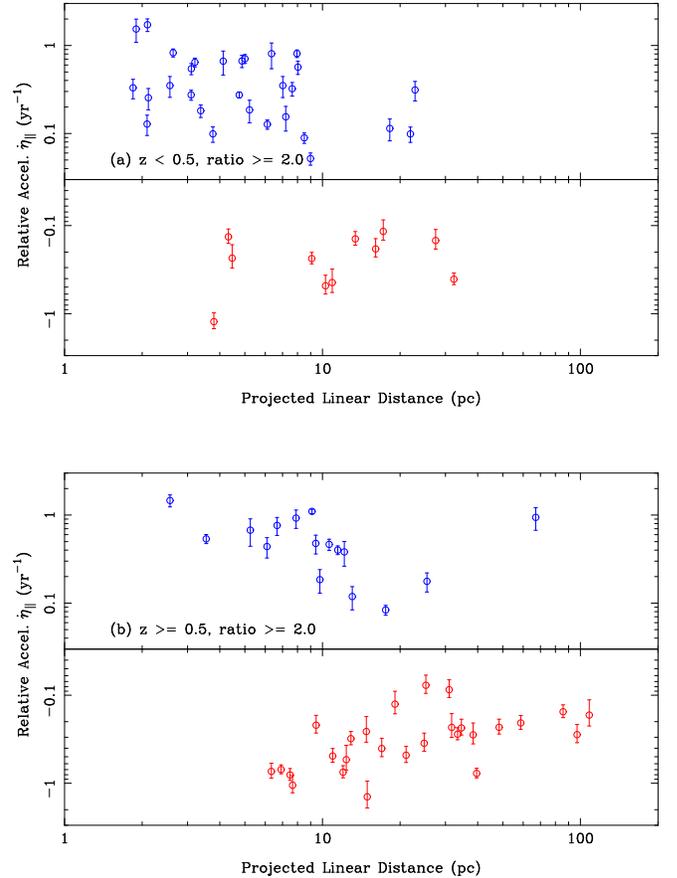}
\figcaption{\label{f:accel_v_pc_z_ratio}
Relative parallel accelerations, $\dot{\eta}_\parallel$ for all jet features with $\geq 3\sigma$ 
parallel accelerations and with parallel to perpendicular acceleration ratios $\geq 2.0$. 
Accelerations are plotted against average projected linear distance from the core 
position in parsecs.  Sources with redshifts of $z < 0.5$ are plotted in panel (a),
and sources with redshifts $z \geq 0.5$ are plotted in panel (b).  In
panel (a), five features from NGC 1052 at projected linear distances $< 1.0$ pc are not
plotted for clarity; four of these have positive acceleration, and one has negative. 
}
\end{figure}

We note that this inability to measure small {\em relative} accelerations
at small angular distances may affect our interpretation of the magnitude
of the acceleration at small distances. Panel (c) of Figure \ref{f:R_dist_accel_plot} 
plots relative accelerations versus projected linear distance, and we have
marked with solid circles those features with $\langle R\rangle \geq 2.0$ mas
to help assess the degree to which features at small projected linear distance
are dominated by measurements made at small angular distance. There is a
thorough mix of features at all projected
distances out to $\simeq 15$ pc. In our subsequent discussion of the {\em magnitude} of relative 
acceleration versus projected linear distance, we will try to mitigate any bias 
due to this effect by also averaging all measured acceleration values, even if they
fall below the $3\sigma$ limit; however, the primary results discussed below
are related to the {\em sign} of the parallel acceleration as a function of
distance, and those results are unaffected.

As found in Paper VII, there is no apparent relationship between
the sign of the acceleration and {\em angular} distance.  We detect
no difference between the angular distance distributions of positive
and negative parallel accelerations with a K$-$S test ($P=0.32$, $N_+=66$, $N_-=57$).  
However, there is clearly a relationship with projected {\em linear} distance,
plotted in panel (c) of Figure \ref{f:R_dist_accel_plot}. 
Jet features are more likely to show positive parallel acceleration
(speeding up) near the base of the jet and negative parallel 
acceleration (slowing down) at larger linear distances from the core. 
A K$-$S test shows a probability $P=6\times10^{-6}$ $(N_+=66$, $N_-=57)$ that the distances
of the positive and negative accelerations are drawn from the same
distribution. The mean projected distances for the positive and negative accelerations
are $9.0$ pc and $22.5$ pc respectively.  It is important to note that for a 
parsec-scale flux-density limited sample, such as MOJAVE, the range of angles to
the line of sight isn't large (with the exception of the handful of radio galaxies
at low redshift); therefore, any intrinsic trends with distance are not 
likely to be smeared out by differing jet orientations \citep{LM97}.

While we argue above and in \S{\ref{s:discuss}} that 
{\em on average} intrinsic changes in Lorentz factor make a larger
contribution to the observed parallel accelerations than
changes in their angle to the line of sight, we do not
know for any individual feature which type of intrinsic
change makes the larger
contribution.  Observed perpendicular accelerations do
give us an indication; however, as we expect that 
large parallel accelerations due solely to jet bending
would be typically accompanied by large perpendicular
accelerations as it is unlikely the bend will be
entirely along our line of sight and changes outside of
our line of sight are greatly magnified due to projection. 
Thus we expect that if we plot only the parallel accelerations 
for jet features with a large parallel/perpendicular 
acceleration ratio, we would be more likely to see the 
effect of Lorentz factor changes than changes to the 
line of sight.  

Figure \ref{f:accel_v_pc_ratio} plots parallel acceleration
versus projected linear distance for all the features 
with a parallel/perpendicular ratio $\geq 2.0$ in panel (a)
and all of those with a ratio $< 2.0$ in panel (b).  
Our expectation is confirmed, as we clearly see a
strong relationship between sign of the parallel 
acceleration and projected linear distance in panel (a)
but not in panel (b). For the jet features with a
parallel/perpendicular ratio $\geq 2.0$, a K$-$S test 
shows a probability $P=2\times10^{-6}$ $(N_+ = 49, N_-=39)$ that the distances
of the positive and negative accelerations are drawn from the same
distribution. The mean distances for the positive and negative 
accelerations are $8.3$ pc and $25.1$ pc respectively. For
jet features with a parallel/perpendicular ratio $< 2.0$, a K$-$S
test detects no significant difference in their distance 
distributions ($P=0.57$, $N_+=17$, $N_-=18$) with means of $11.0$
and $16.9$ pc for positive and negative accelerations, respectively. 

Figure \ref{f:accel_v_pc_z_ratio} is a plot of the jet
features in Figure \ref{f:accel_v_pc_ratio}(a) broken into
low redshift, $z<0.5$ (panel a), and high redshift, $z\geq0.5$
(panel b). The difference between positive and negative
accelerations shows up in the same manner for both low and 
high redshift sources and is significant in both cases: $P=0.003$ $(N_+=33, N_-=12)$ 
and $P=0.01$ $(N_+=16, N_-=27)$, respectively.

\begin{figure}
\includegraphics[scale=0.45,angle=0,totalheight=0.45\textheight]{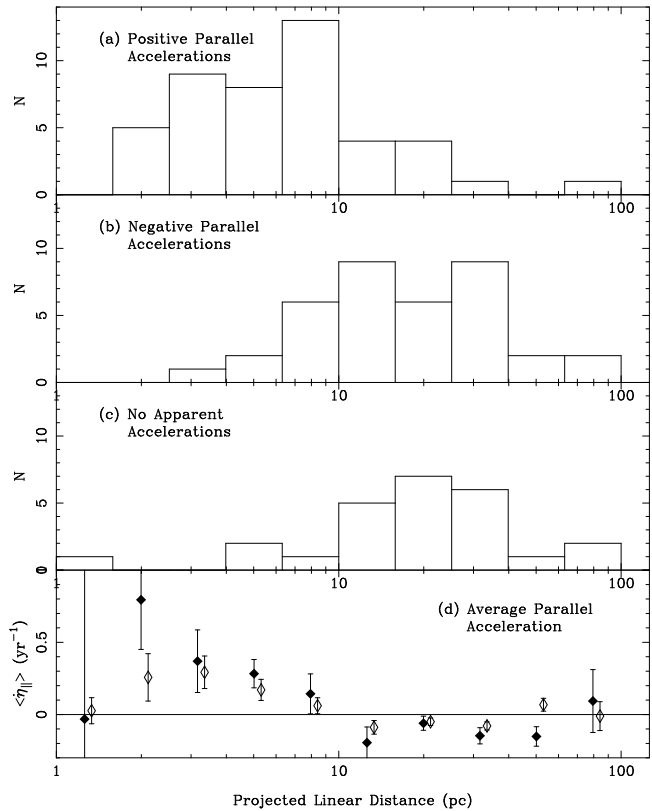}
\figcaption{\label{f:accel_v_pc_hist}
Histograms of projected linear distance for jet features with positive parallel
accelerations (panel (a)), negative parallel accelerations (panel (b)),
and no apparent acceleration (panel (c)).  
Panels (a) and (b) only include features with parallel accelerations significant 
at the $\geq 3\sigma$ level and with parallel to perpendicular acceleration 
ratios $\geq 2.0$.  Panel (c) selects features with no significant acceleration
and robustly small parallel and perpendicular accelerations defined by 
$|\dot{\eta}_{\parallel,\perp}|+2\sigma \leq 0.1$.  
Note that seven features from NGC 1052 at projected 
linear distances $< 1.0$ pc are not plotted in the above three panels for clarity; four of these have 
positive acceleration, one has negative acceleration, and two have no apparent
acceleration.
In Panel (d), solid diamonds 
show the unweighted average of the relative parallel accelerations in each 
projected distance bin for the features 
plotted in the above panels. Open diamonds give the average acceleration in
each projected distance bin using all features in Table 1, regardless of
significance level of the acceleration measurement.
}
\end{figure}

To further evaluate the relationship between parallel
acceleration and projected distance, histograms of 
the distances in Figure \ref{f:accel_v_pc_ratio}(a) are given in
the first two panels of Figure \ref{f:accel_v_pc_hist}.  Panel
(c) of Figure \ref{f:accel_v_pc_hist} is a histogram of projected
distances for features with  no apparent parallel or perpendicular acceleration.  The
distance histogram of features with no apparent acceleration is similar
to the distance histogram for features with negative parallel
acceleration, but both clearly differ from the histogram of features
with positive parallel acceleration.  The bottom panel, (d), of
Figure \ref{f:accel_v_pc_hist} shows the average relative parallel 
acceleration in each distance bin for the features plotted in the 
three histograms above.
The average relative acceleration decreases with distance from 
large positive values at very short projected distances to 
progressively smaller averages, until the average acceleration
switches sign beyond $10$ pc, where it takes on modest values
up to $-0.2$ yr$^{-1}$. 

\subsection{Perpendicular Acceleration vs. Proper Motion Mis-alignment}

Here we explore the relationship between observed 
perpendicular acceleration and apparent mis-alignment
between the proper motion vector direction, $\phi$,
and the mean structural position angle of 
the jet feature, $\langle\vartheta\rangle$.  Jet
features that have a significant mis-alignment between
their proper motion vector and structural position angle 
have non-ballistic motion, and therefore, we
expect that they have experienced perpendicular
accelerations that have changed their trajectory
since being ejected from the base of the jet.

\begin{figure}
\includegraphics[scale=0.45,angle=-90,totalheight=0.5\textheight]{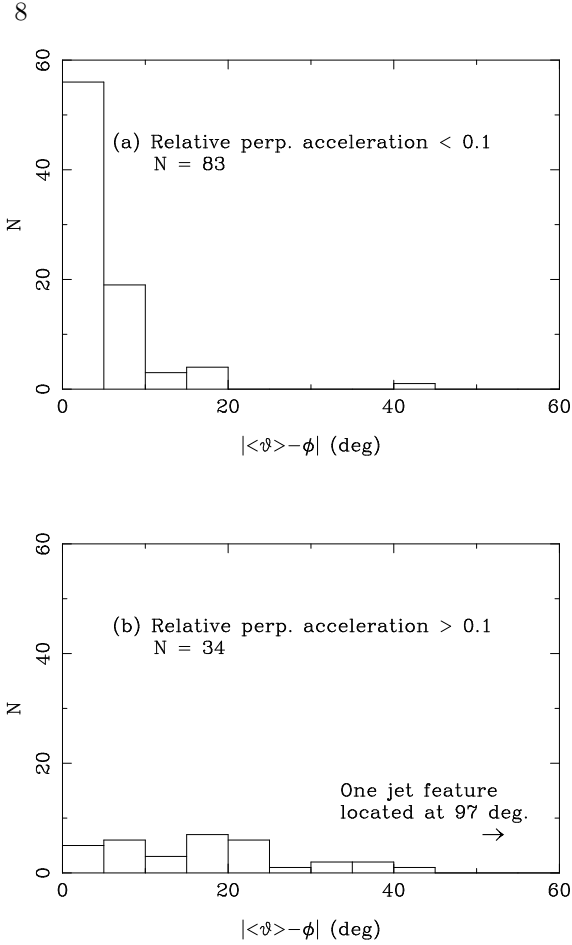}
\figcaption{\label{f:nonradial_motion_dualhist}
Histograms of the magnitude of the misalignment angle, $|\langle\vartheta\rangle-\phi|$, for jet features 
with small perpendicular accelerations (panel (a)) of $|\dot{\eta}_\perp| + 2\sigma < 0.1$ and those 
with large perpendicular acceleration (panel (b)) of $|\dot{\eta}_\perp| - 2\sigma > 0.1$.  
}
\end{figure}

Figure \ref{f:nonradial_motion_dualhist} shows 
histograms of the magnitude of the proper motion 
mis-alignment with respect to the mean structural 
position angle of the jet feature for two types
of jet features.  Panel (a) includes features
with small perpendicular accelerations, $|\dot{\eta}_\perp| +2\sigma \leq 0.1$,
and Panel (b) includes features with large perpendicular
accelerations, $|\dot{\eta}_\perp|-2\sigma > 0.1$.  The
two distributions clearly differ ($P < 10^{-8}$) with 
large perpendicular accelerations much more likely to be linked 
with large mis-alignments; however, it is noteworthy
that some jet features with small perpendicular
accelerations do have large mis-alignment angles, perhaps 
because they experienced acceleration prior to
our monitoring period.

\begin{figure}
\includegraphics[scale=0.4,angle=-90,totalheight=0.26\textheight]{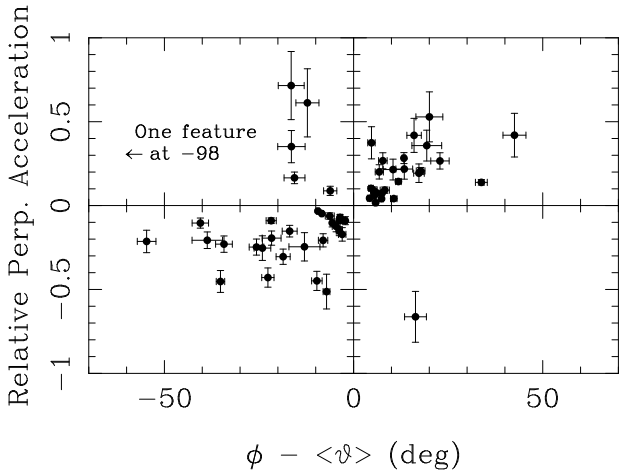}
\figcaption{\label{f:perp_accel_v_offset}
Plot of relative perpendicular acceleration, $\dot{\eta}_\perp$, vs. non-radial motion 
misalignment angle for jet features with $\geq 3\sigma$ significant perpendicular acceleration and 
non-radial motion. Of the 59 features plotted, 52 have acceleration in the correct
direction to have caused the observed misalignment (i.e., are located in the 1st and 3rd
quadrants).
}
\end{figure}

Figure \ref{f:perp_accel_v_offset} plots relative
perpendicular acceleration versus proper
motion mis-alignment angle\footnote{Note the
convention $\phi-\langle\vartheta\rangle$ is the opposite of
that chosen in Figure 10 in Paper VII; however, 
this sign definition simplifies the discussion.} 
for all jet features with both significant perpendicular
acceleration {\em and} significant non-radial
motion.  Positive 
perpendicular acceleration will tend to increase
$\phi$ over time, so the observed accelerations are
overwhelmingly ($52$ out of $59$ features) in the 
correct direction to have caused the observed mis-aligned
motion.

Knowing that observed perpendicular accelerations cause
the observed mis-aligned motions does not tell us why
the features are changing direction to become non-radial.  
\citet{KL04} showed that the non-radial motion of jet
features tends to be in the direction of the downstream 
emission, suggesting that jets follow pre-established
channels of flow.  This result was confirmed in Paper VII.
Here we examine the relationship with {\em upstream} emission 
as defined by the mean inner jet position angle.  In 
Paper X, we calculated the mean inner jet position angle
by averaging all CLEAN components in the radio map between radii of 0.15
and 1.0 mas from the optically thick core feature near the base of the jet.  
This calculation
was performed for all epochs, resulting in a circular mean and  
standard deviation for the inner jet position angle
for each source.  We note that, as described in Paper X, 
mean inner jet position angles were not available for 
all sources in our sample due to insufficient time coverage or number
of epochs, core identification uncertainty, counter-jet
emission, or highly curved jet structure within 1 mas of 
the core.  Note that the circular standard deviation in the jet
position angle is mainly due to the variability in jet position
angle over time as newly emitted jet features emerge at 
different position angles (Paper X).

\begin{figure}
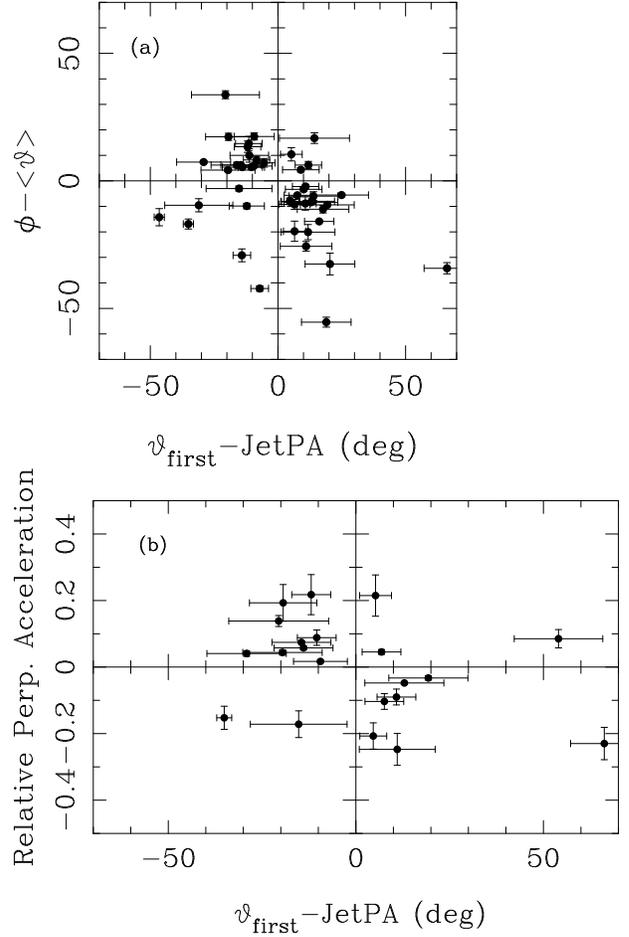

\includegraphics[scale=0.4,angle=-90,totalheight=0.26\textheight]{fig9a.eps}\\
\includegraphics[scale=0.4,angle=-90,totalheight=0.26\textheight]{fig9b.eps}
\figcaption{\label{f:initial_angle}
Plots of non-radial motion misalignment angle, $\phi-\langle\vartheta\rangle$, and relative 
perpendicular acceleration, $\dot{\eta}_\perp$, versus initial structural
angle misalignment, $\vartheta_{first}-$JetPA for jet features at $\langle R\rangle \geq 2.0$ mas 
with $\geq 3\sigma$ non-radial motion (panel (a)) or perpendicular accelerations (panel (b)).  
Here $\vartheta_{first}$ is the
structural position angle of the feature in the first epoch it was observed, and JetPA 
is the mean position angle of the inner jet over all epochs as defined in Paper X.
Note that the $x$-axis error bars indicate the standard deviation of inner jet position
angle over in all epochs that source was observed.  Only jet features with 
an initial structural position angle outside the standard deviation of inner jet 
position angle measurements are plotted. In general, the observed non-radial motions
and perpendicular accelerations are in the correct direction to move the jet feature
toward the mean position angle of the inner jet ($35$ of the $46$ plotted
non-radial motions, $16$ of the $21$ of the plotted perpendicular accelerations).
}
\end{figure}

In Figure \ref{f:initial_angle} we plot both non-radial 
motion mis-alignment angle and relative perpendicular 
acceleration against the offset between the first epoch
structural position angle of a jet feature, $\vartheta_{first}$,
and the mean of the inner jet position angle, JetPA.  Our key
question is whether jet features that start out with
a large offset between their structural position angle
and the mean inner jet position angle experience 
kinematics that tend to bring them into better 
alignment with the inner jet position angle.
We define a `large offset' to be a first
epoch structural position angle that is more than one standard deviation
from the mean inner jet position angle, where the circular standard
deviation is computed for the set of inner jet position angles 
across all epochs in which that source was observed.
It is also important that we only consider jet features
with mean angular distance $\langle R\rangle \geq 2.0$ mas
to avoid the possibility that the jet feature we are
studying contributes significantly to the mean of the
inner jet position angle\footnote{Including those features
with $\langle R\rangle < 2.0$ mas, strengthens the statistical
relationships we find, but this may be an artifact of those features 
moving toward their own mean position.}. 

Both the observed non-radial motions and perpendicular 
accelerations have a strong tendency to be in the correct
direction to move the jet feature toward the direction
of the mean inner jet position angle, indicating some
sort of collimating effect prevents features from moving
along a ballistic trajectory.  The odds of at
least $35$ out of $46$ non-radial motions and $16$ out of $21$
perpendicular accelerations having this tendency by pure chance are
$P = 0.00027$ and $P = 0.013$, respectively. 

\begin{figure}
\includegraphics[scale=0.4,angle=0,totalheight=0.52\textheight]{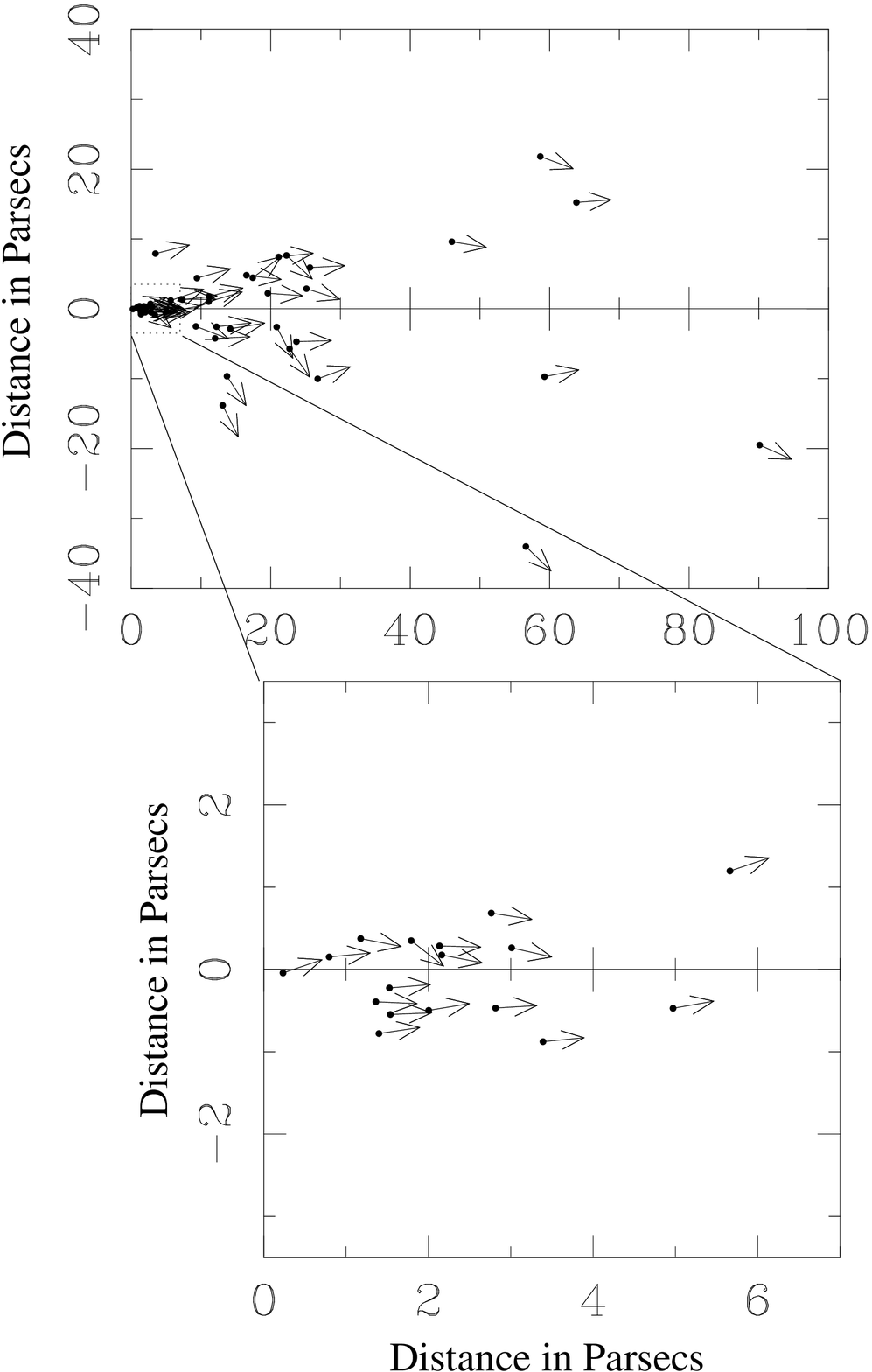}
\figcaption{\label{f:collim_vect}
Locations of the 46 jet features plotted in Figure 9(a) in their first epoch of 
observation with an arrow showing the direction of their mean velocity vector. Positions of 
jet features from multiple sources are plotted together with projected linear distance from
the core, given in parsecs.  The locations of the features for each source are rotated so
that the $x$-axis represents the mean inner jet position angle.  The inset shows
the details of the cluster of features at short projected distances.
}
\end{figure}

As a diagnostic of this relationship, Figure \ref{f:collim_vect}
plots each of the jet features in Figure \ref{f:initial_angle}(a)
at their first epoch position with an arrow indicating the
direction of their velocity vector.  Jet features
from different sources are included on the plot by using
projected linear distance in parsecs and rotating the 
position angles to place the mean inner jet position angle 
of each source along the $x$-axis in the figure.  The
inset of Figure \ref{f:collim_vect} clearly shows that 
the tendency for jet features to experience motions that 
move them toward better alignment with the mean inner
jet position angle is very strong for features at small
projected linear distance. Almost all (16 out of 17) 
features plotted in the inset have this tendency; however, for
jet features first observed at larger distances, the tendency
is much weaker (19 out 29) and more consistent with pure chance.
At these larger projected distances, jets may have already experienced
bends, and those jet features may be following the bends as 
indicated in \citet{KL04} and Paper VII.

\section{Discussion}
\label{s:discuss}

We find that apparent accelerations in the motion of jet
features are common, with approximately half of the motions
we studied showing significant accelerations 
parallel and/or perpendicular to their velocity
vector.  We find at least one significantly
accelerating feature in three-quarters of the blazar jets in our
sample.  Parallel accelerations, indicating
changes in apparent speed, are a factor of two times
larger on average than perpendicular accelerations associated with
changes in direction.  This confirms our result
from Paper VII, also verified by \citet{P12}, that parallel
accelerations cannot be the result of jet bending alone.
  
As demonstrated in Paper VII and summarized in \S{2}, changes
in jet direction can change the apparent speed of jet
features; however, due to projection effects, we expect to 
see even larger changes in velocity direction on the sky (i.e. 
perpendicular accelerations).  From the beaming characteristics
of a typical parsec-scale flux-density limited sample, Paper VII showed
that parallel accelerations due to bending alone should be only 60\% of
the magnitude of the accompanying perpendicular accelerations.
Here we see almost the opposite ratio, with parallel accelerations
nearly a factor of two times the magnitude of perpendicular accelerations.
While for any given jet feature, we cannot confidently ascribe 
an apparent parallel acceleration to a change in Lorentz factor; we
conclude that, on average, changes in the Lorentz factors of jet 
features dominate the observed parallel accelerations.

\subsection{Lorentz Factor vs. Jet Distance}

Paper VII reported an overall tendency for jet features
with positive parallel acceleration to be at smaller projected 
linear distances than jet features with negative parallel 
acceleration, with a fuzzy break between positive and negative 
acceleration occurring at $\sim 15$ pc.  \citet{P12} confirmed
the tendency for a dominance of positive accelerations at 
short projected distances $< 15$ pc, but their smaller sample,
focused on compact sources, 
did not allow them to measure the acceleration of more than a 
few features beyond this distance. In a earlier, smaller study of 
$\gamma$-ray blazars at 43 GHz, \citet{J05} reported accelerations in 
nine of the fifteen jets they studied, and these
tended to be positive parallel accelerations near the base of jet, 
although they could not determine whether the observed accelerations 
were due to changes in intrinsic speed or direction of the 
jet features. 

Figures 3 through 6 explore this relationship and we clearly 
confirm the tendency for jet features with positive parallel
acceleration (speeding up) to appear at shorter projected
linear distances than jet features with negative parallel
acceleration (slowing down), with much better statistics than
we achieved in Paper VII. It is noteworthy that Figure 3(b)
shows no relationship if the accelerations are plotted against {\em
angular} distance, the direct observable in our maps and 
motion fitting, thus ruling out simple observational biases.
Projected {\em linear} distance depends on source redshift,
so we also investigated this relationship in both low and
high redshift groups (Figure 5) and found it holds
in both separately.  

Figure 4 tests this relationship as a function of 
parallel/perpendicular acceleration ratio.  As discussed above, 
accelerations of jet features with high parallel/perpendicular 
ratios are expected to be more likely to be due to changes 
in Lorentz factor.  If this is true {\em and} if the relationship
with jet distance is due to changes in Lorentz factor, we would 
expect the relationship with jet distance to be stronger for 
jet features with a high parallel/perpendicular ratio, and that is 
precisely what we found.

Finally, Figure 6 compares these high parallel/perpendicular
ratio accelerating features to each other and to a set of
features that we can be confident have little, if any, 
apparent acceleration.  Jet features with positive parallel 
acceleration are clearly at smaller projected distance than
negatively accelerating features or features with no
apparent acceleration; however, features with negative
apparent acceleration do not appear to differ in their
distance distribution from jet features with no 
apparent acceleration.  The final panel of Figure 6 shows 
the average acceleration in 
each distance bin and clearly indicates a decrease from 
strong positive acceleration at short distances to 
slightly negative acceleration at distances $\gtrsim 10$ pc.

Equation 4 allows us to estimate the relative change in Lorentz
factor in the galaxy frame under the assumption that all of the
observed relative acceleration is due to Lorentz factor changes. 
If $\Gamma$ is sufficiently greater than unity, then $\beta\simeq 1$
and $\dot{\Gamma}/\Gamma \simeq \dot{\eta}_\parallel/\delta^2$.
The average positive relative accelerations range from $0.1$ to $1.0$ yr$^{-1}$
for projected distances $\lesssim 10$ pc in Figure 6(d).  Assuming
Doppler factors of order $\delta \sim 10$, the implied positive
acceleration in the core region falls in the range 
$\dot{\Gamma}/\Gamma \simeq 10^{-3}$ to $10^{-2}$ per year in the
galaxy frame.  Due to our detection of several rapidly accelerating
features at small projected distances, the rates of change estimated 
above are up to an order of magnitude larger than those estimated 
in Paper VII and found by \citet{P12} in their sample.

It is important to note that while rates of
intrinsic change in range $\dot{\Gamma}/\Gamma \simeq 10^{-3}$ to $10^{-2}$ per year 
may seem modest, they are acting over very long periods
of time in the frame of the AGN host galaxy where a decade of observed
time in our frame may span centuries.  A jet feature at a projected distance of $10$ pc
actually lies $\sim10^2$ pc from the nucleus, assuming a typical order of 
magnitude de-projection factor in a highly beamed jet.  This large
de-projected distance indicates that the 
process of acceleration of features in the jet flow may span hundreds of
light years in the source frame.  As we show in \S{\ref{s:speed_dist}},
the observed accelerations are in many cases sufficient to double
or triple the apparent speed of a moving feature during our
observations, suggesting a corresponding change in the Lorentz factor of
the same order.

Beyond projected distances of $\sim 10$ pc, jet features are
more likely to show negative acceleration or no detectable acceleration 
with average observed accelerations in the range $0.0$ to $-0.2$
yr$^{-1}$ corresponding to source frame Lorentz factor decreases
in the range $\dot{\Gamma}/\Gamma \sim -10^{-3}$ per year.
As discussed in Paper VII, Lorentz factor decreases of this magnitude 
would completely decelerate jets before they reach scales of tens to
hundreds of kiloparsecs; however, Lorentz factor decreases brought
about by interaction with the inter-stellar medium would be
expected to decrease in magnitude with distance from the center
of the galaxy.

\subsubsection{Apparent Speed vs. Distance}
\label{s:speed_dist}

Measurements of apparent speeds of different jet features in the
same source have been used as evidence that jets show positive 
acceleration on parsec scales, with features at small projected 
distances appearing to be slower than features found further away 
\citep{Kr98,Cot99,H01,PMF07,B08,P12}.  Paper X reported 
a general trend of increasing apparent speed with distance down the
jet for radio galaxies and BL Lac objects. Plots of apparent speed 
versus distance in M87 by \citet{MSB13} and \citet{AND14} show increasing apparent 
speed out to projected linear distances of $\sim70$ pc at HST-1 and decreasing
apparent speed at larger distances. Note that the de-projection 
factor for M87 is only a factor of $2-4$ assuming a viewing angle
in the range $15-25^\circ$ \citep{AAA09}, so the de-projected
transition from increasing speed to constant or decreasing speed occurs
 at $\sim 10^2$ pc, consistent with our estimates in the previous
section. 

In this section we discuss plots of apparent speed versus projected 
linear distance which include {\em speed profiles} for our 
accelerating jet features. A speed profile is simply 
a plot of our best fit accelerating kinematic model to 
show the apparent speed of the feature as a function of projected linear distance
down the jet.  We show these speed profiles only for the inner 60\% of
our observed time range to avoid inadequate extrapolation of the instantaneous speeds deduced from
our kinematic models beyond the point where they are well supported by the data. 

\begin{figure*}
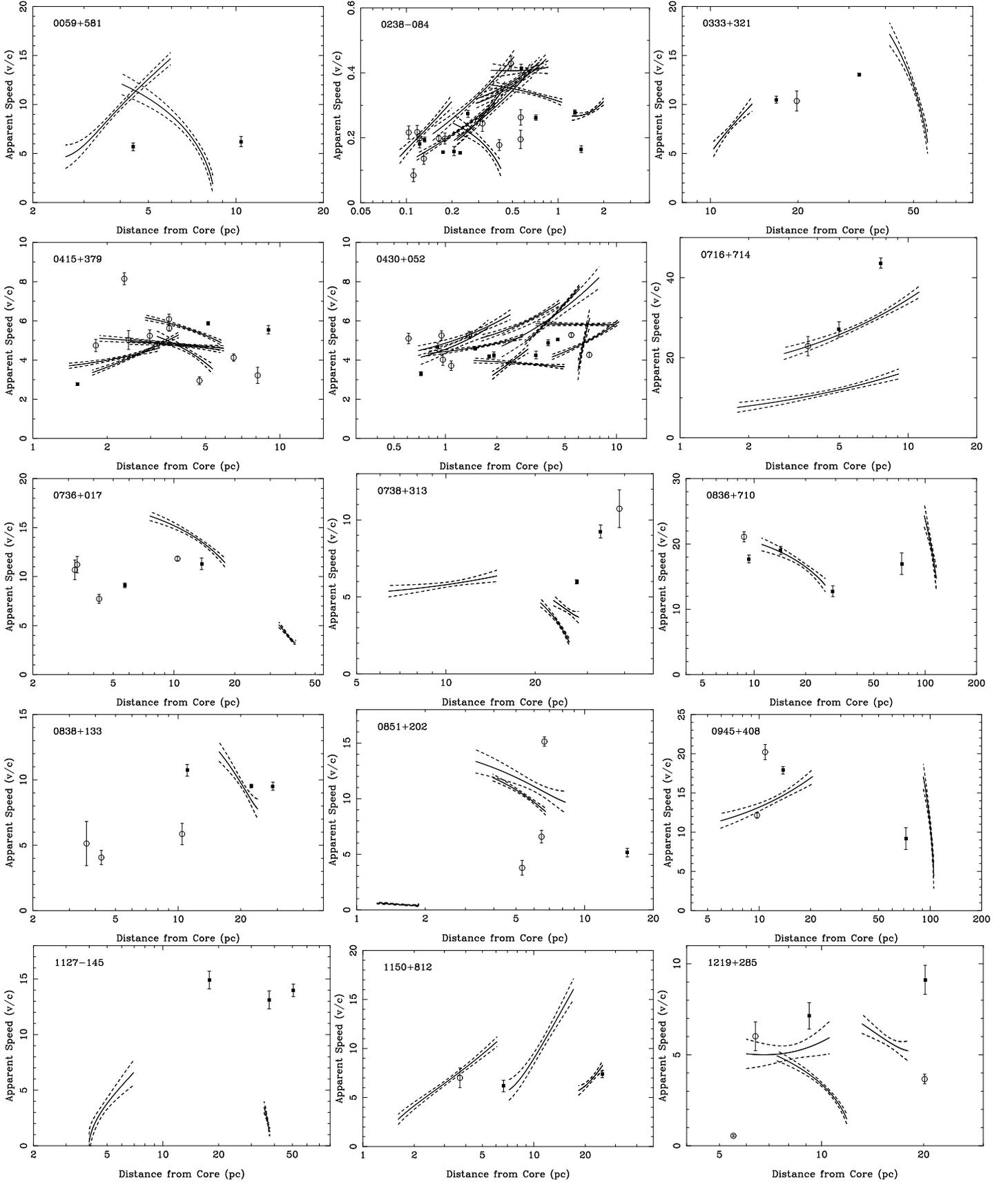

\includegraphics[scale=0.4,angle=-90,totalheight=0.19\textheight]{fig11_0059+581.eps}
\includegraphics[scale=0.4,angle=-90,totalheight=0.19\textheight]{fig11_0238-084.eps}
\includegraphics[scale=0.4,angle=-90,totalheight=0.19\textheight]{fig11_0333+321.eps}\\
\includegraphics[scale=0.4,angle=-90,totalheight=0.19\textheight]{fig11_0415+379.eps}
\includegraphics[scale=0.4,angle=-90,totalheight=0.19\textheight]{fig11_0430+052.eps}
\includegraphics[scale=0.4,angle=-90,totalheight=0.19\textheight]{fig11_0716+714.eps}\\
\includegraphics[scale=0.4,angle=-90,totalheight=0.19\textheight]{fig11_0736+017.eps}
\includegraphics[scale=0.4,angle=-90,totalheight=0.19\textheight]{fig11_0738+313.eps}
\includegraphics[scale=0.4,angle=-90,totalheight=0.19\textheight]{fig11_0836+710.eps}\\
\includegraphics[scale=0.4,angle=-90,totalheight=0.19\textheight]{fig11_0838+133.eps}
\includegraphics[scale=0.4,angle=-90,totalheight=0.19\textheight]{fig11_0851+202.eps}
\includegraphics[scale=0.4,angle=-90,totalheight=0.19\textheight]{fig11_0945+408.eps}\\
\includegraphics[scale=0.4,angle=-90,totalheight=0.19\textheight]{fig11_1127-145.eps}
\includegraphics[scale=0.4,angle=-90,totalheight=0.19\textheight]{fig11_1150+812.eps}
\includegraphics[scale=0.4,angle=-90,totalheight=0.19\textheight]{fig11_1219+285.eps}
\figcaption{\label{f:speed_v_distance}
Plots of apparent speed vs. projected linear distance for all sources with four or more jet features
in Table 1. Features with significant acceleration are plotted as linear projections of their fitted 
motion with one-sigma uncertainties defined by the upper and lower dashed lines. Features
without significant accelerations are plotted as solid squares at the projected location of 
the midpoint of their motion. The apparent proper motion of other robust features from 
Paper X are plotted as open circles.
}
\end{figure*}

Figure \ref{f:speed_v_distance} shows individual source
plots for all $33$ jets which have at least four jet features suitable
for acceleration analysis in this paper. Speed profiles are only
plotted for features showing significant parallel and/or perpendicular
acceleration. Other jet features studied in this paper without significant
acceleration are plotted as
solid squares, and we've included all the remaining robust jet features 
from Paper X as open circles.  Clear increases in apparent speed with
projected linear distances are seen in many of the jets, e.g., 1928$+$738. 
A trend of decreasing
speeds at larger distances is seen in several cases as well. 
These trends are noisy but generally support our conclusions
from the acceleration measurements themselves in the previous section. 

\begin{figure*}
\figurenum{\ref{f:speed_v_distance}}
\includegraphics[scale=0.4,angle=-90,totalheight=0.19\textheight]{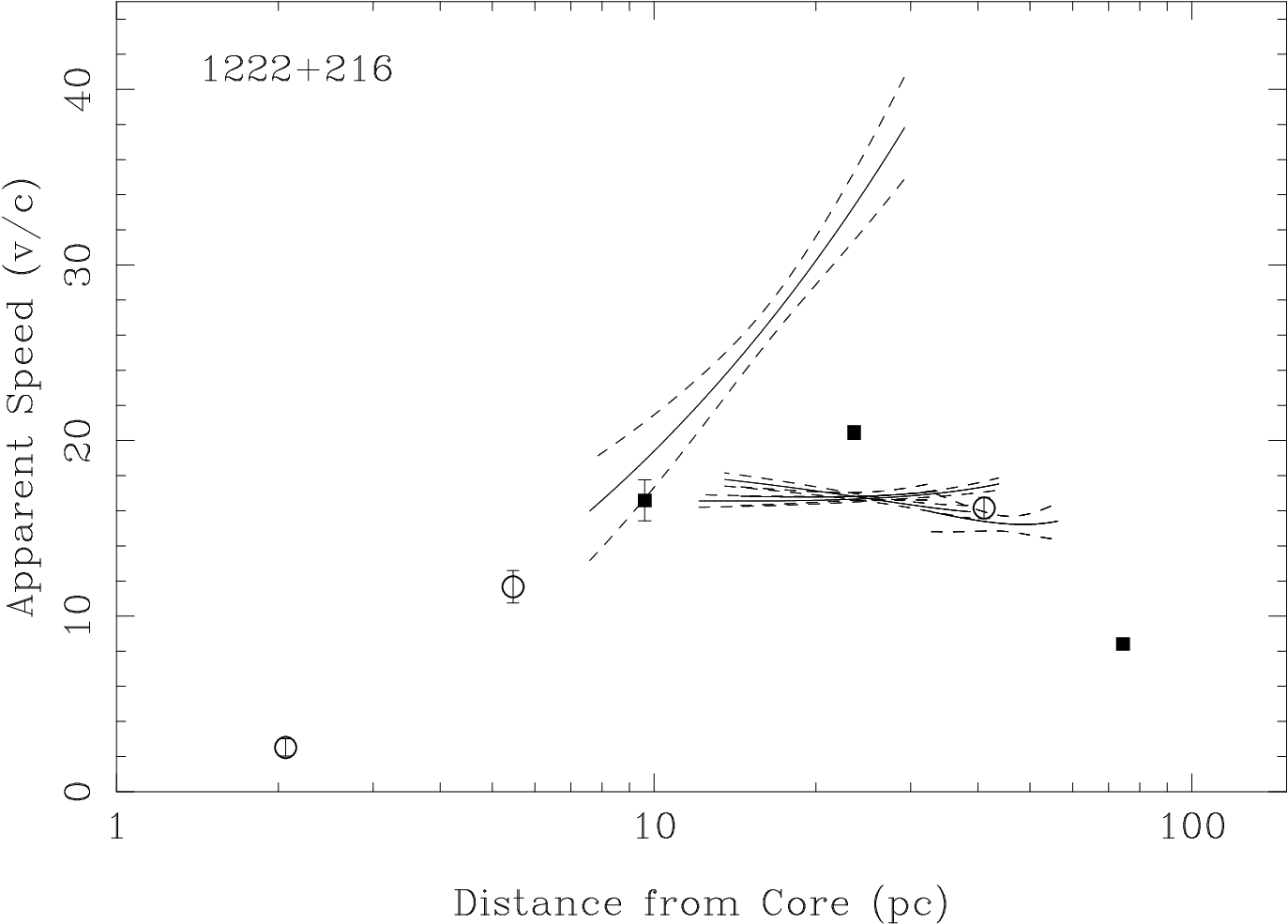}
\includegraphics[scale=0.4,angle=-90,totalheight=0.19\textheight]{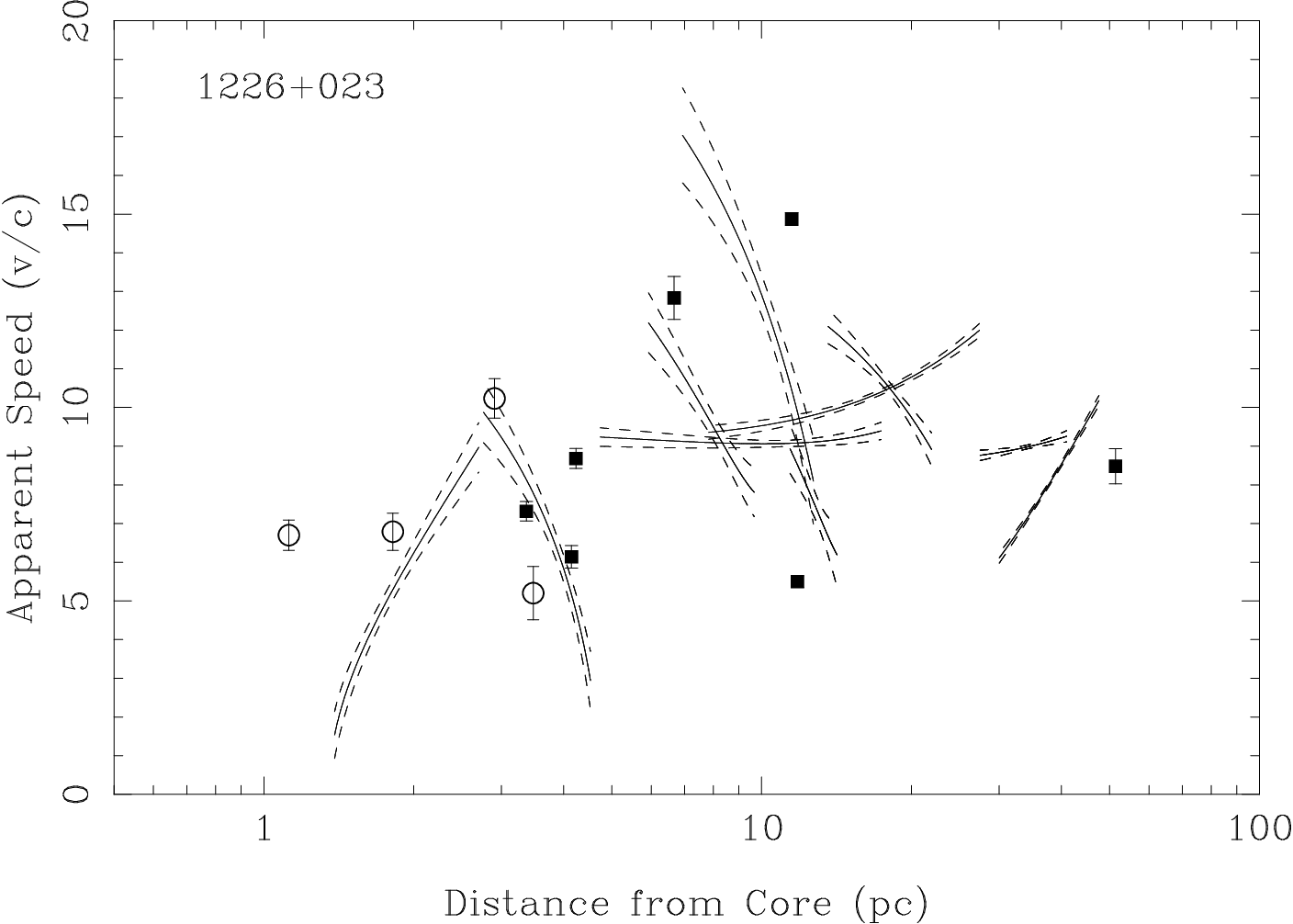}
\includegraphics[scale=0.4,angle=-90,totalheight=0.19\textheight]{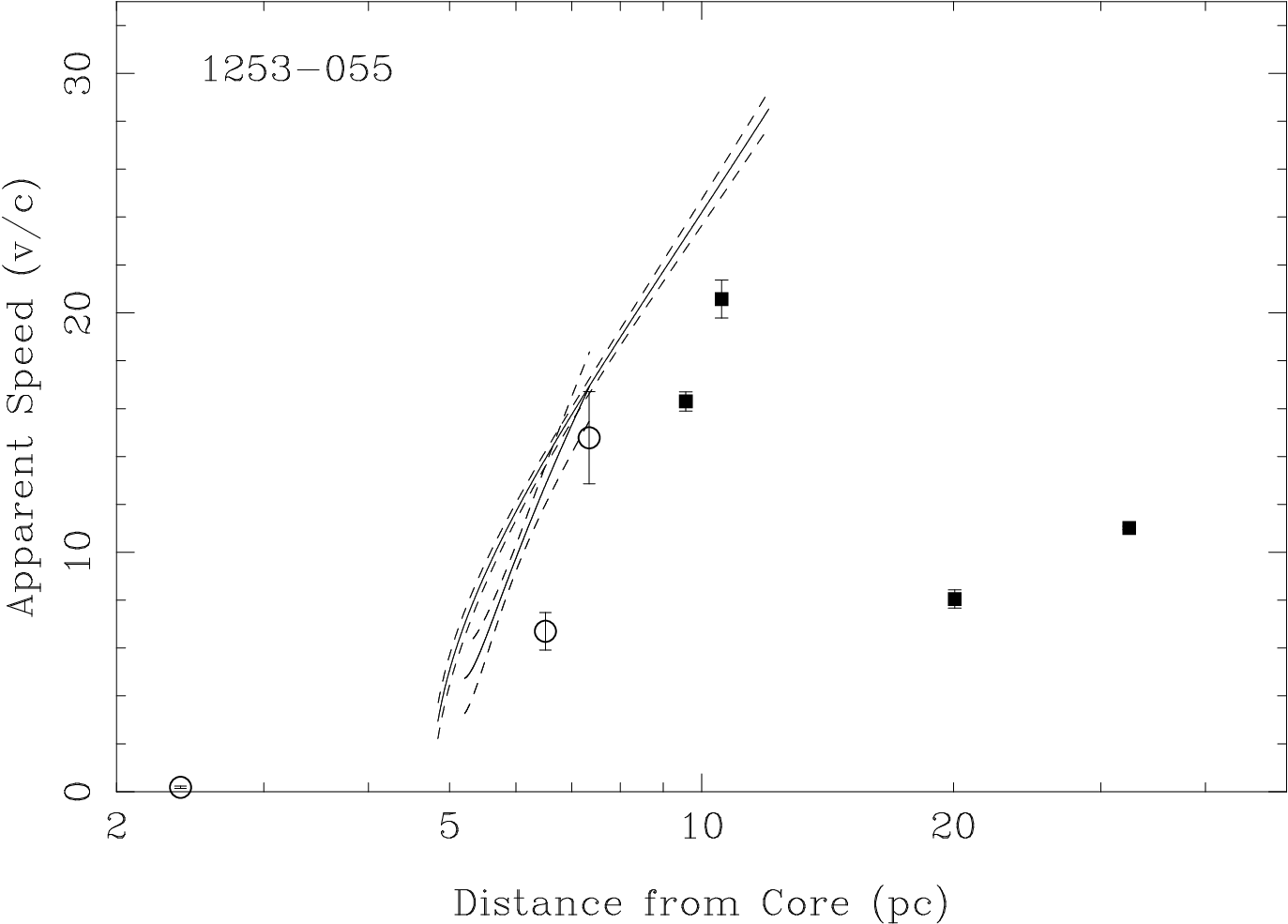}\\
\includegraphics[scale=0.4,angle=-90,totalheight=0.19\textheight]{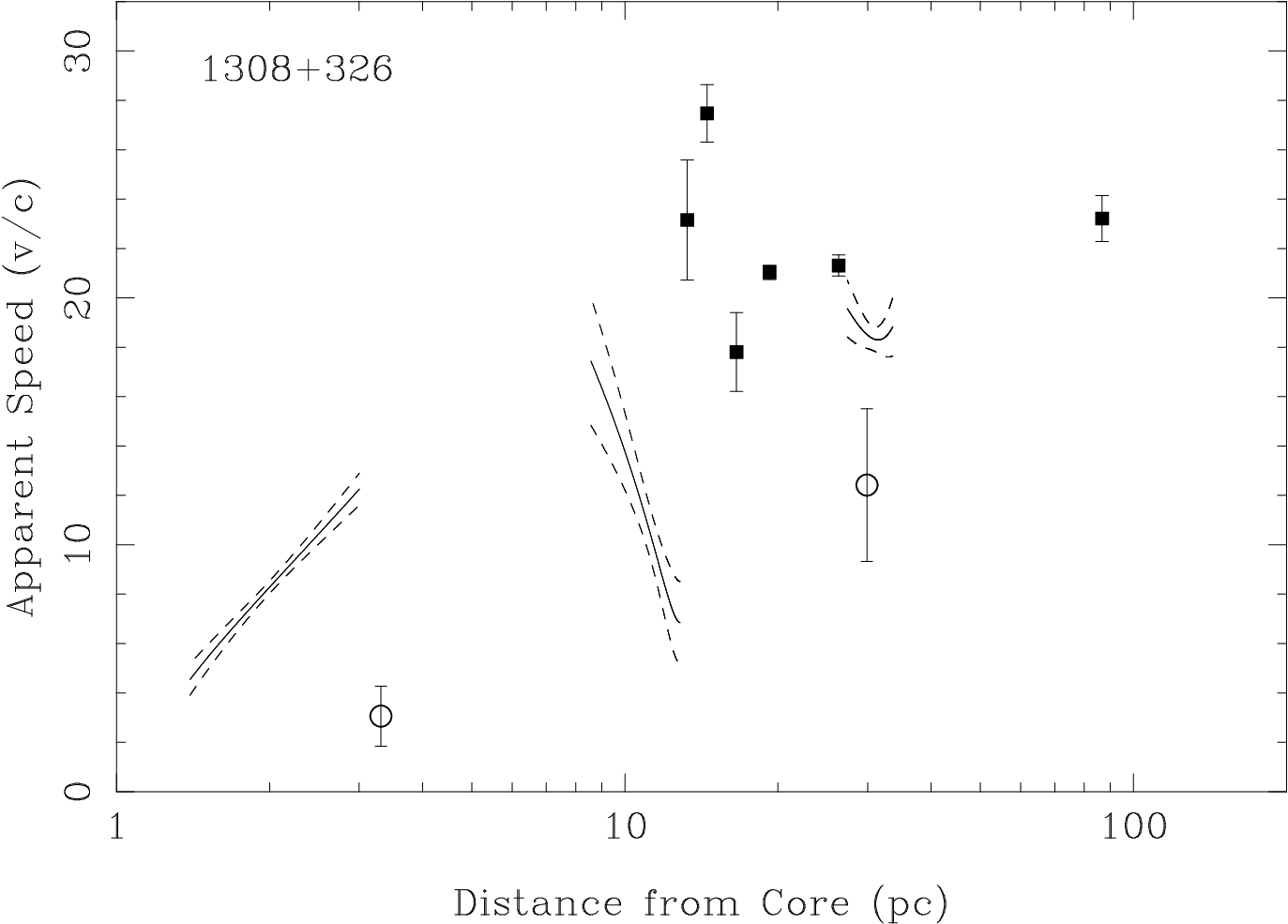}
\includegraphics[scale=0.4,angle=-90,totalheight=0.19\textheight]{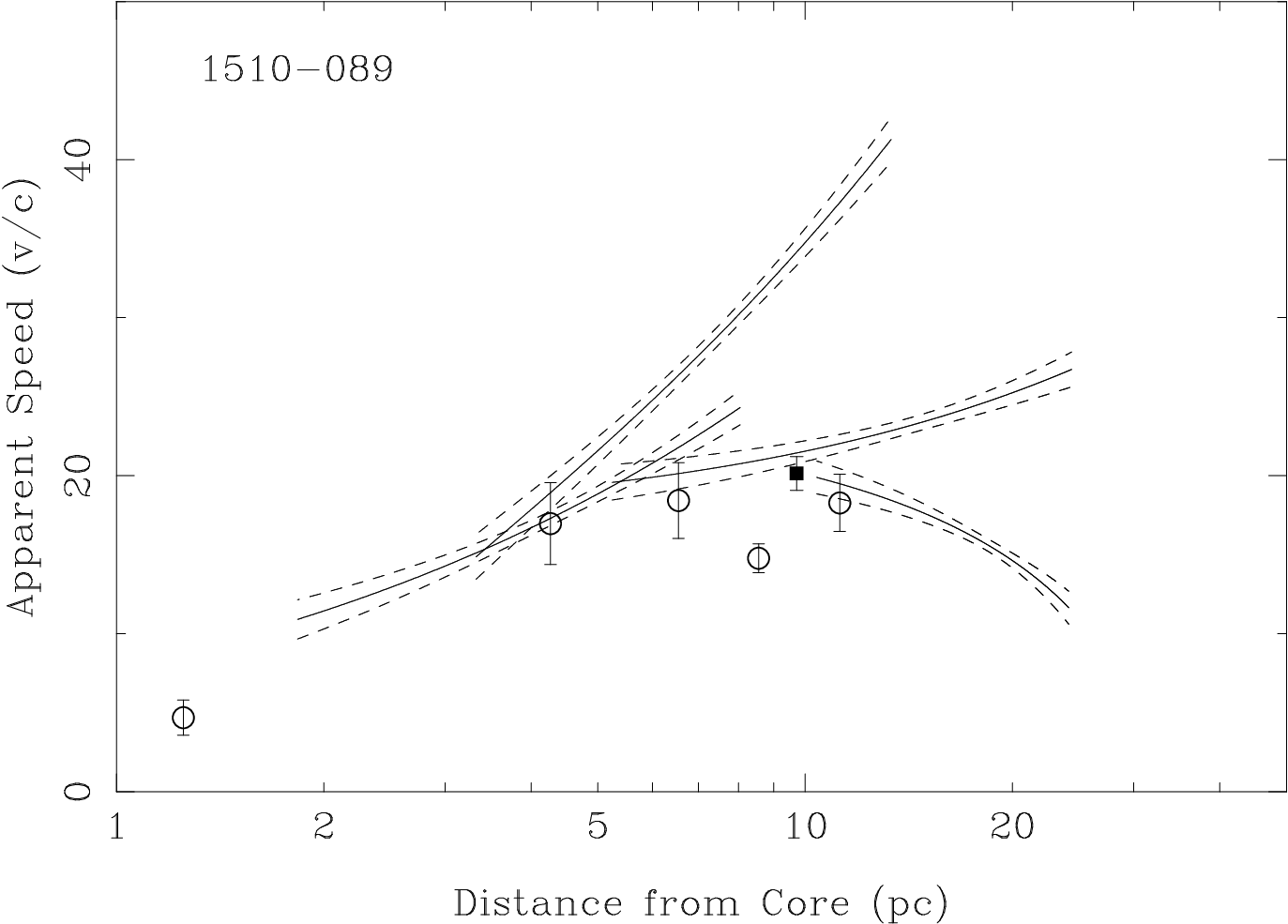}
\includegraphics[scale=0.4,angle=-90,totalheight=0.19\textheight]{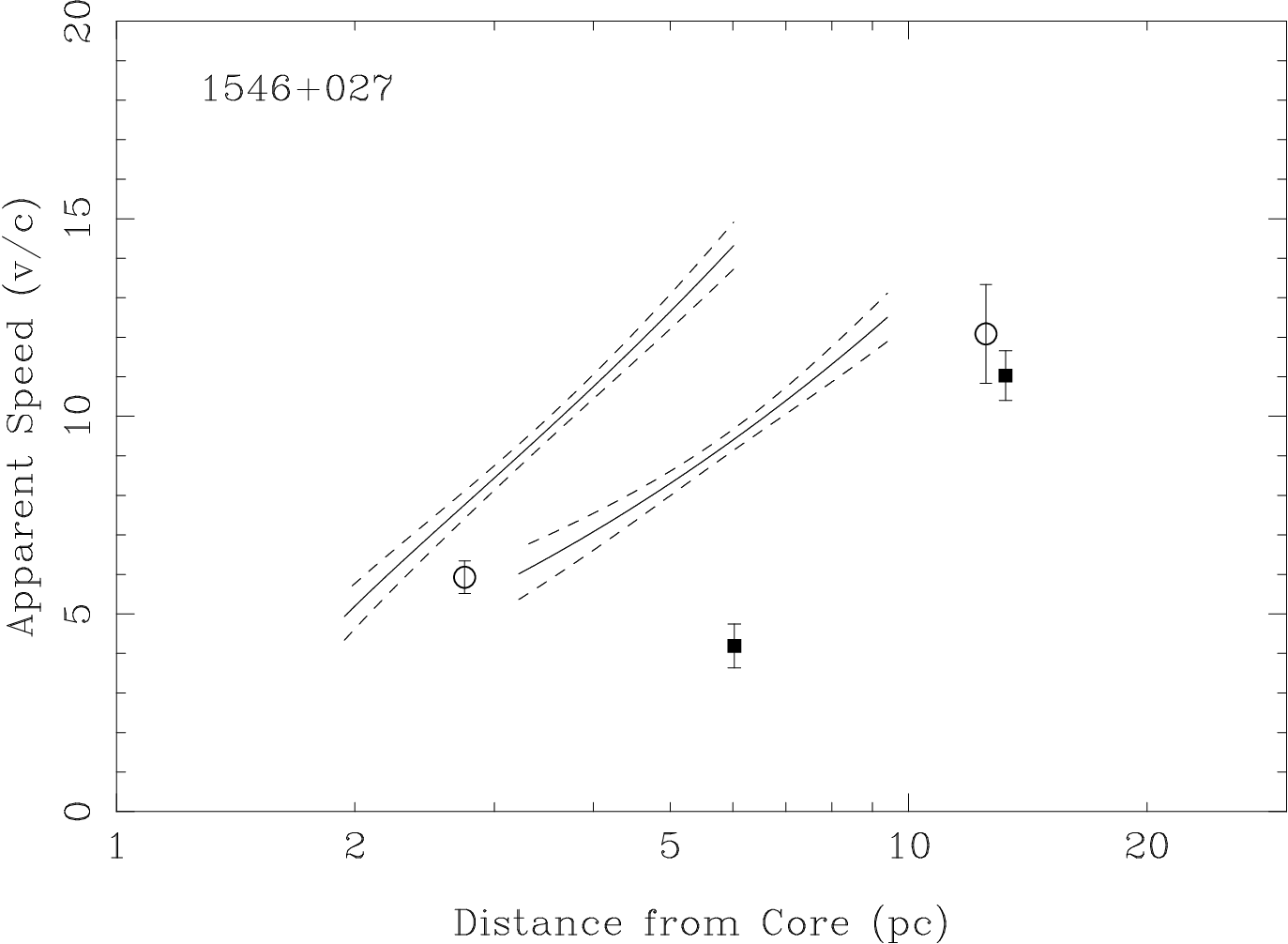}\\
\includegraphics[scale=0.4,angle=-90,totalheight=0.19\textheight]{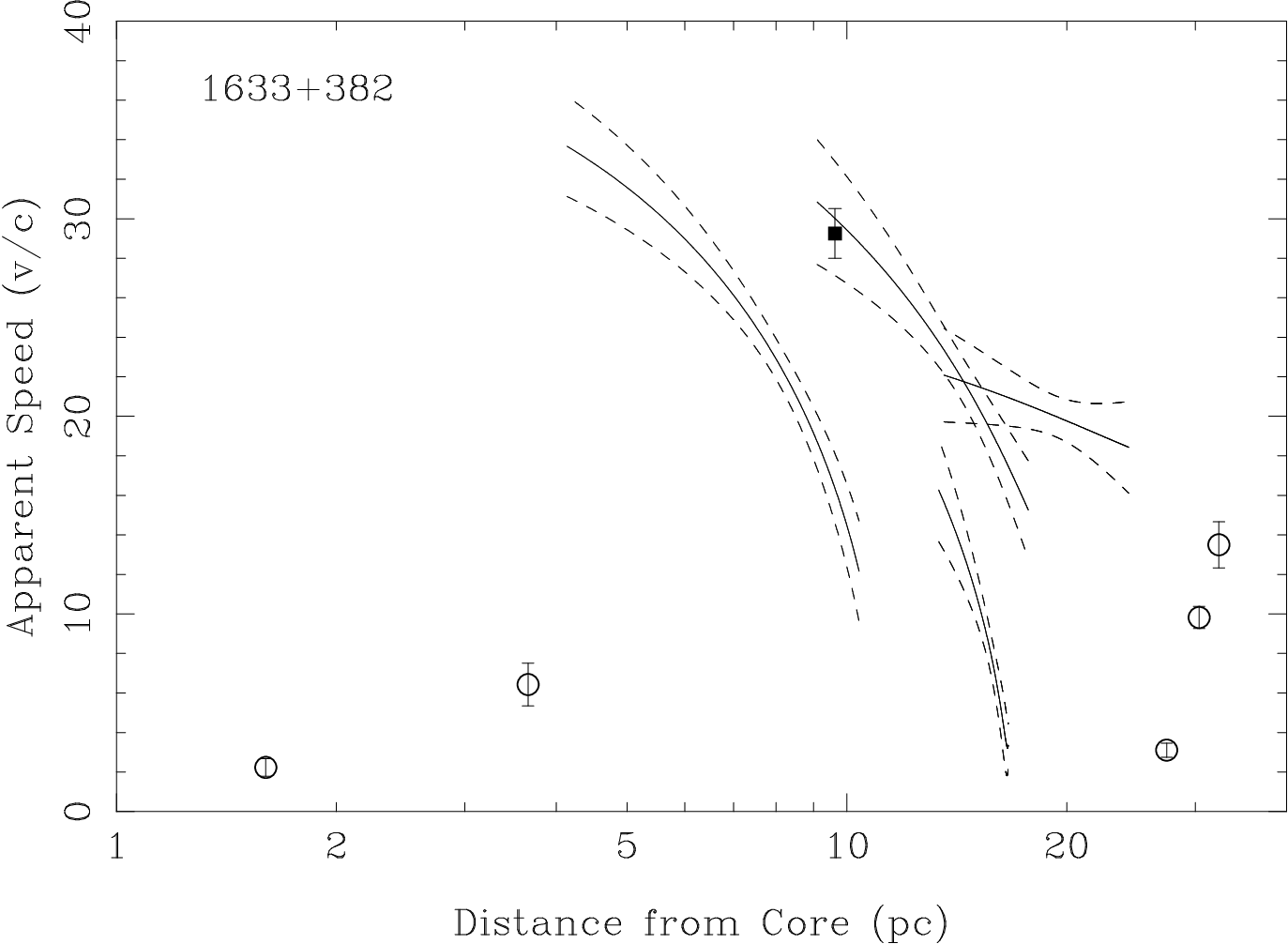}
\includegraphics[scale=0.4,angle=-90,totalheight=0.19\textheight]{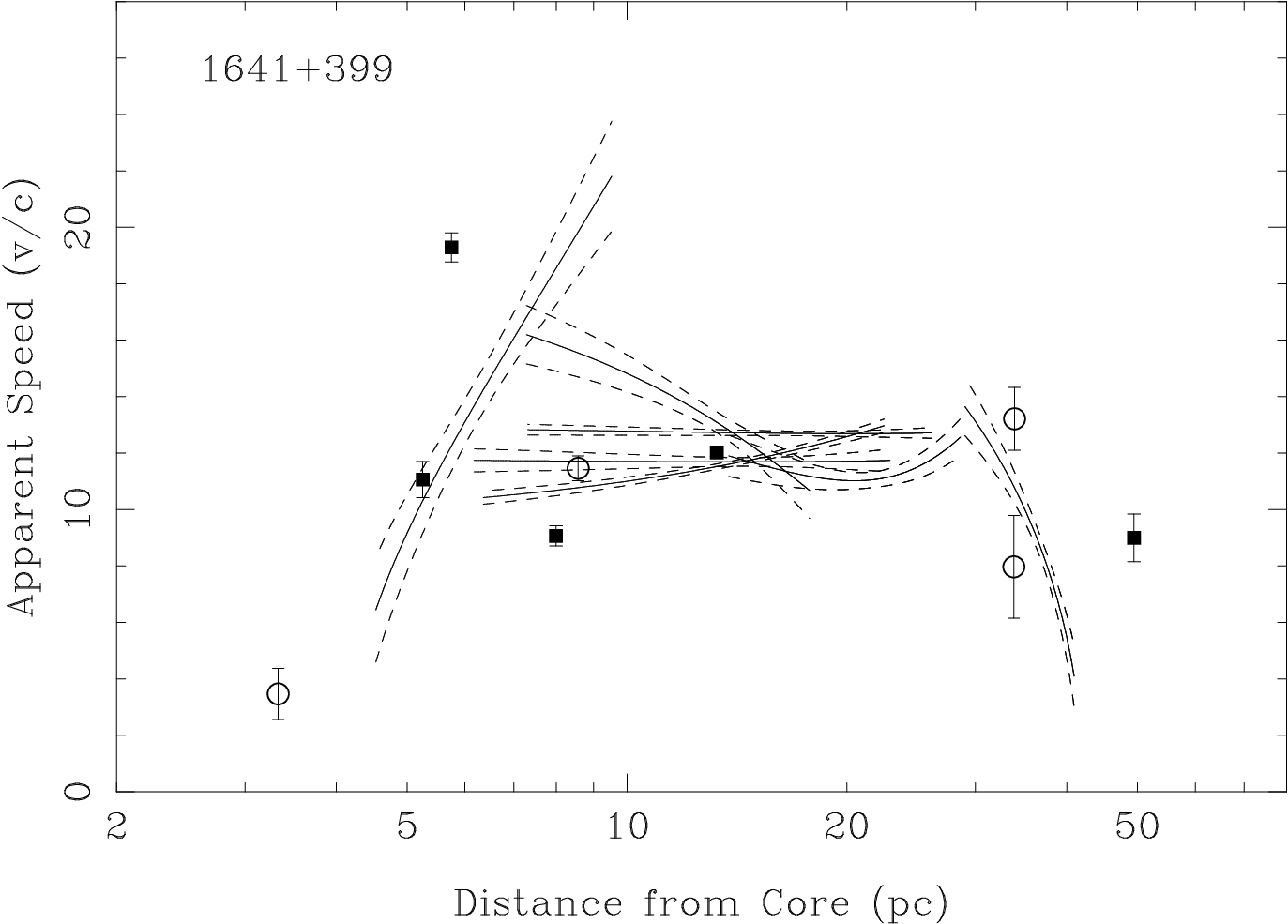}
\includegraphics[scale=0.4,angle=-90,totalheight=0.19\textheight]{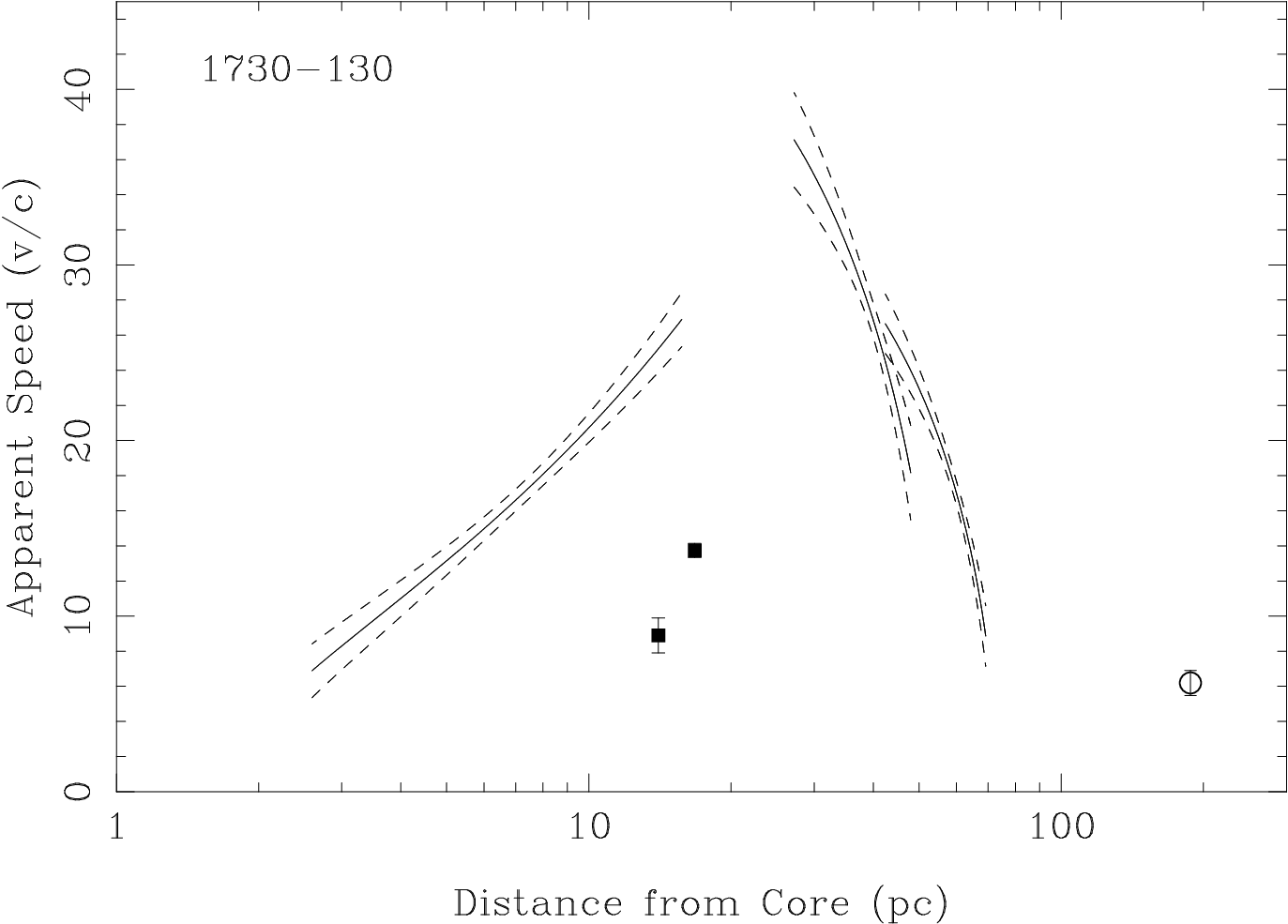}\\
\includegraphics[scale=0.4,angle=-90,totalheight=0.19\textheight]{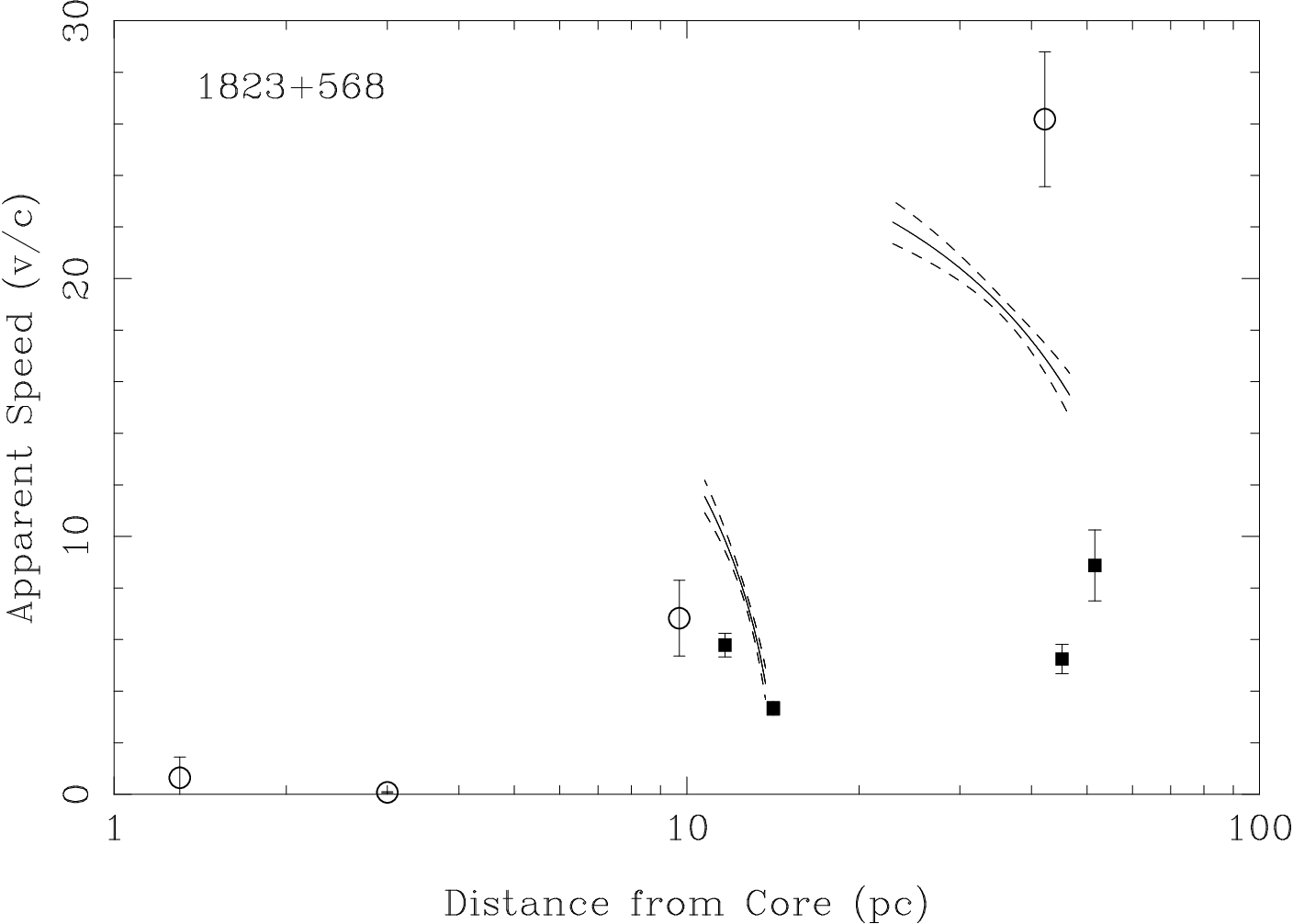}
\includegraphics[scale=0.4,angle=-90,totalheight=0.19\textheight]{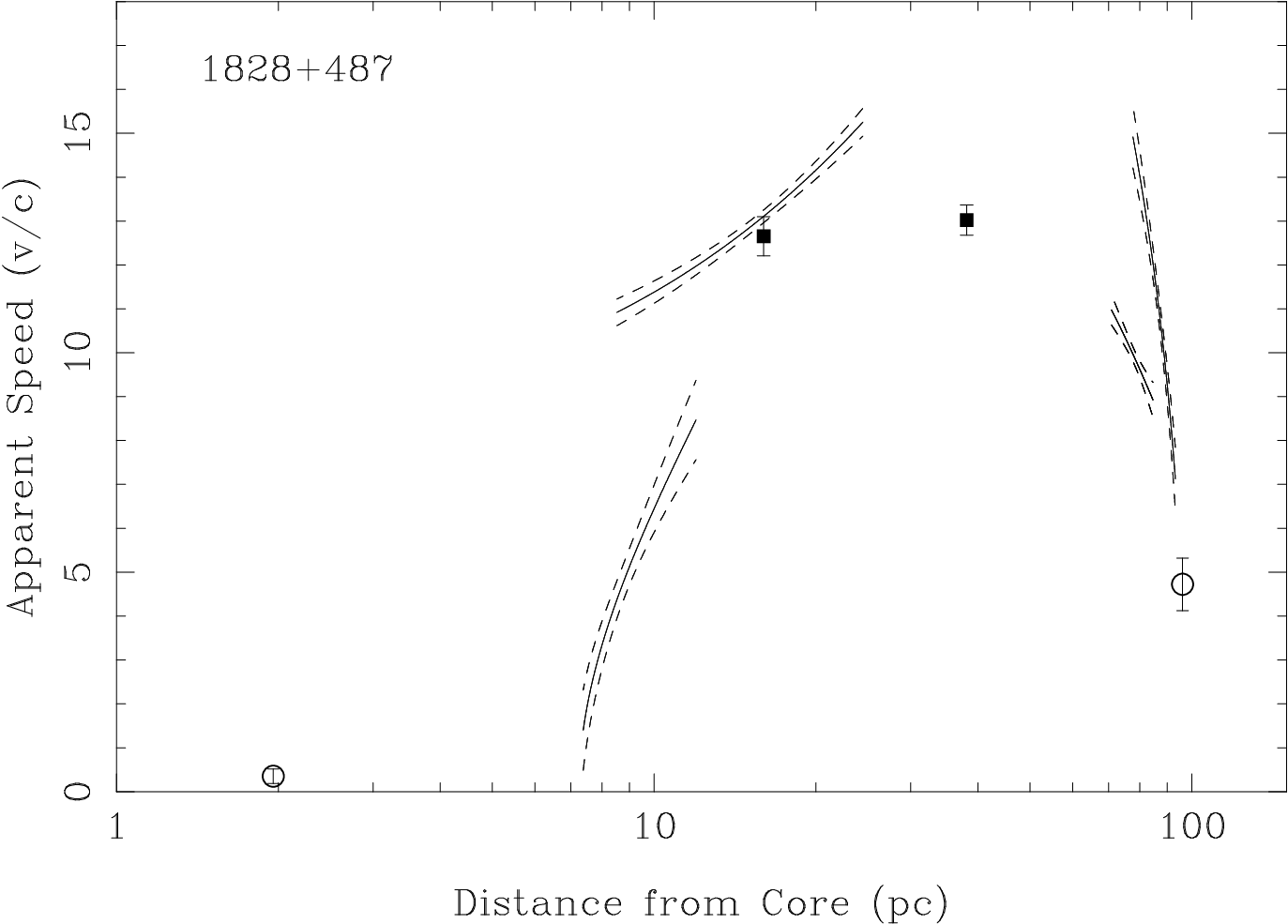}
\includegraphics[scale=0.4,angle=-90,totalheight=0.19\textheight]{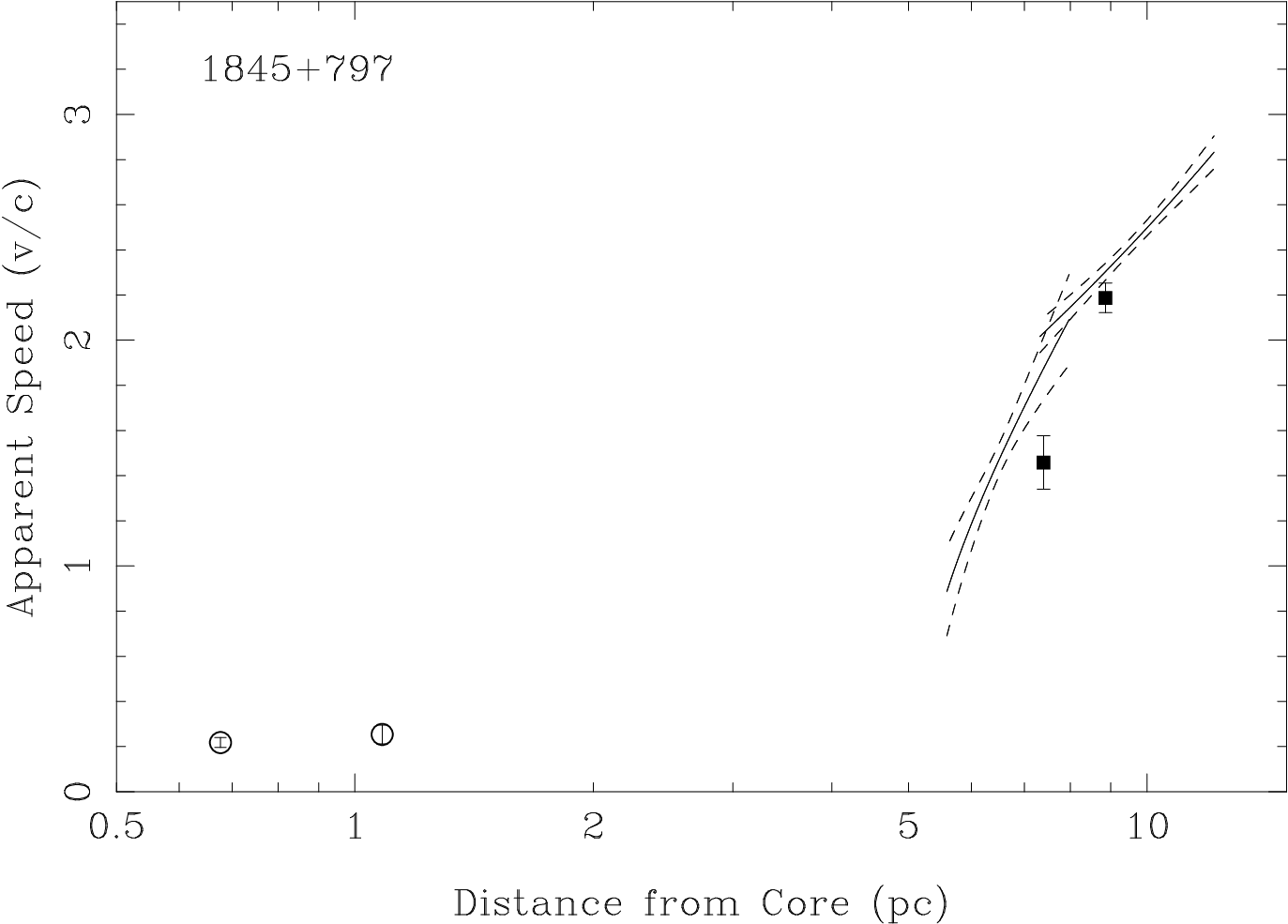}\\
\includegraphics[scale=0.4,angle=-90,totalheight=0.19\textheight]{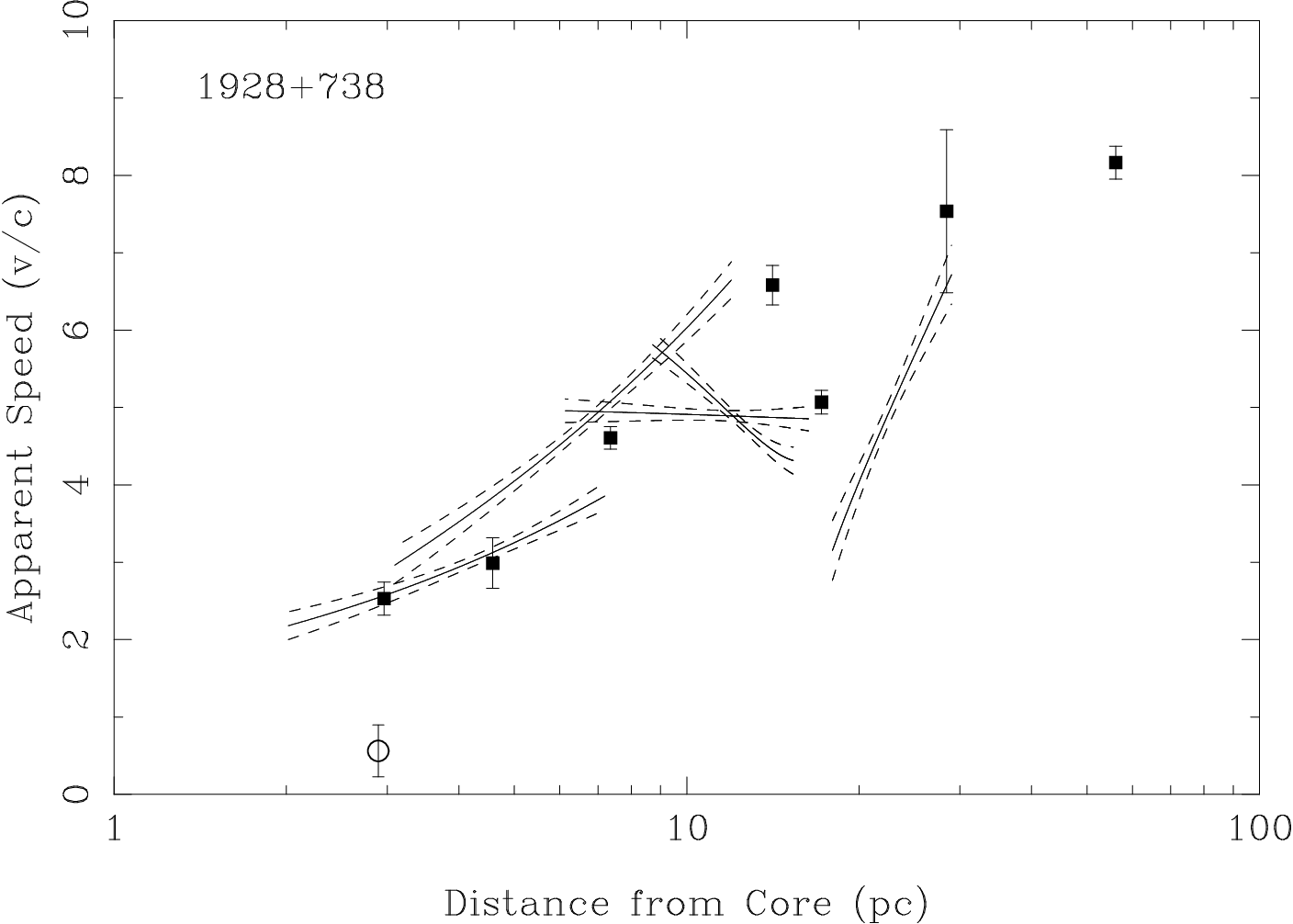}
\includegraphics[scale=0.4,angle=-90,totalheight=0.19\textheight]{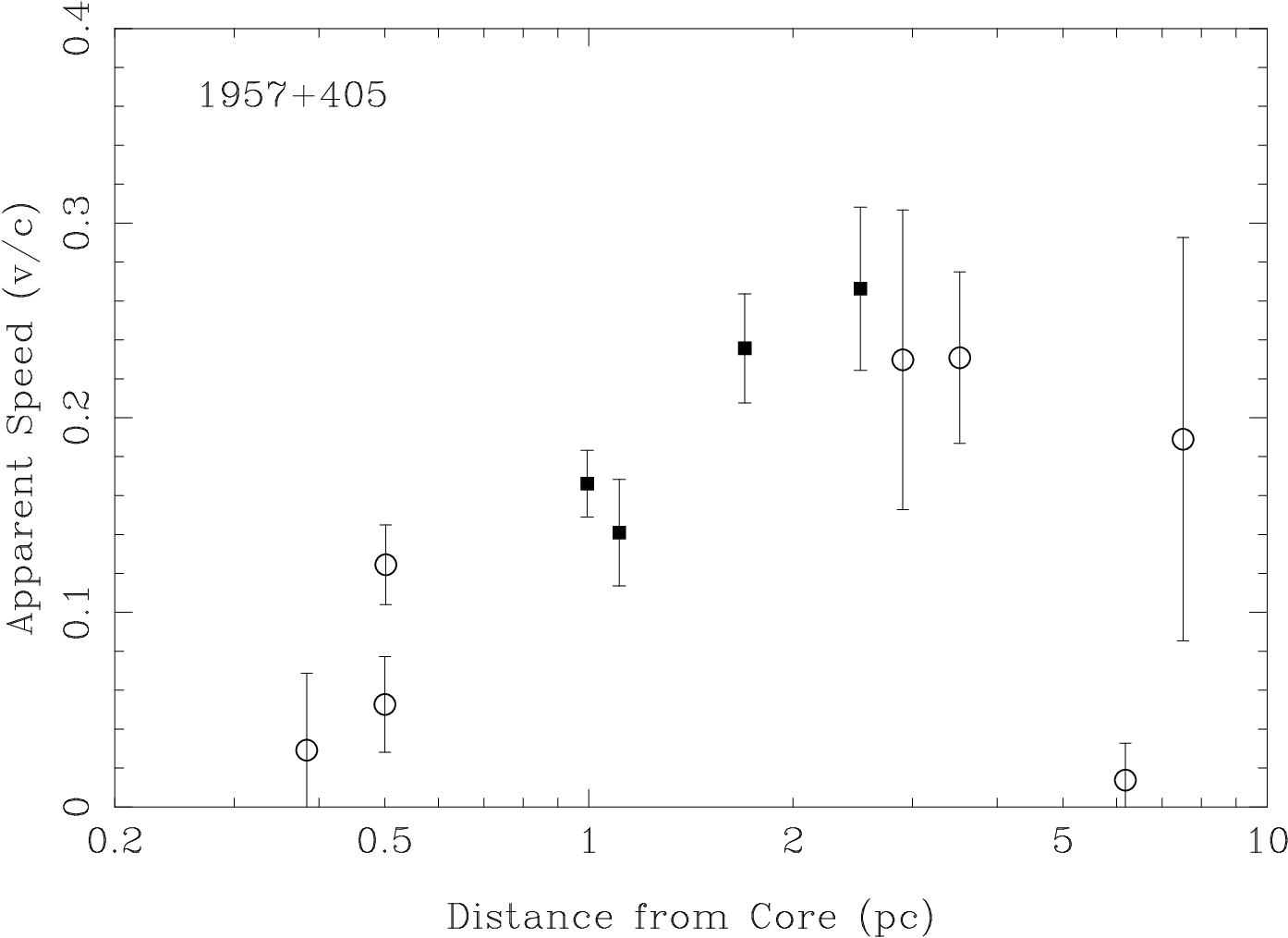}
\includegraphics[scale=0.4,angle=-90,totalheight=0.19\textheight]{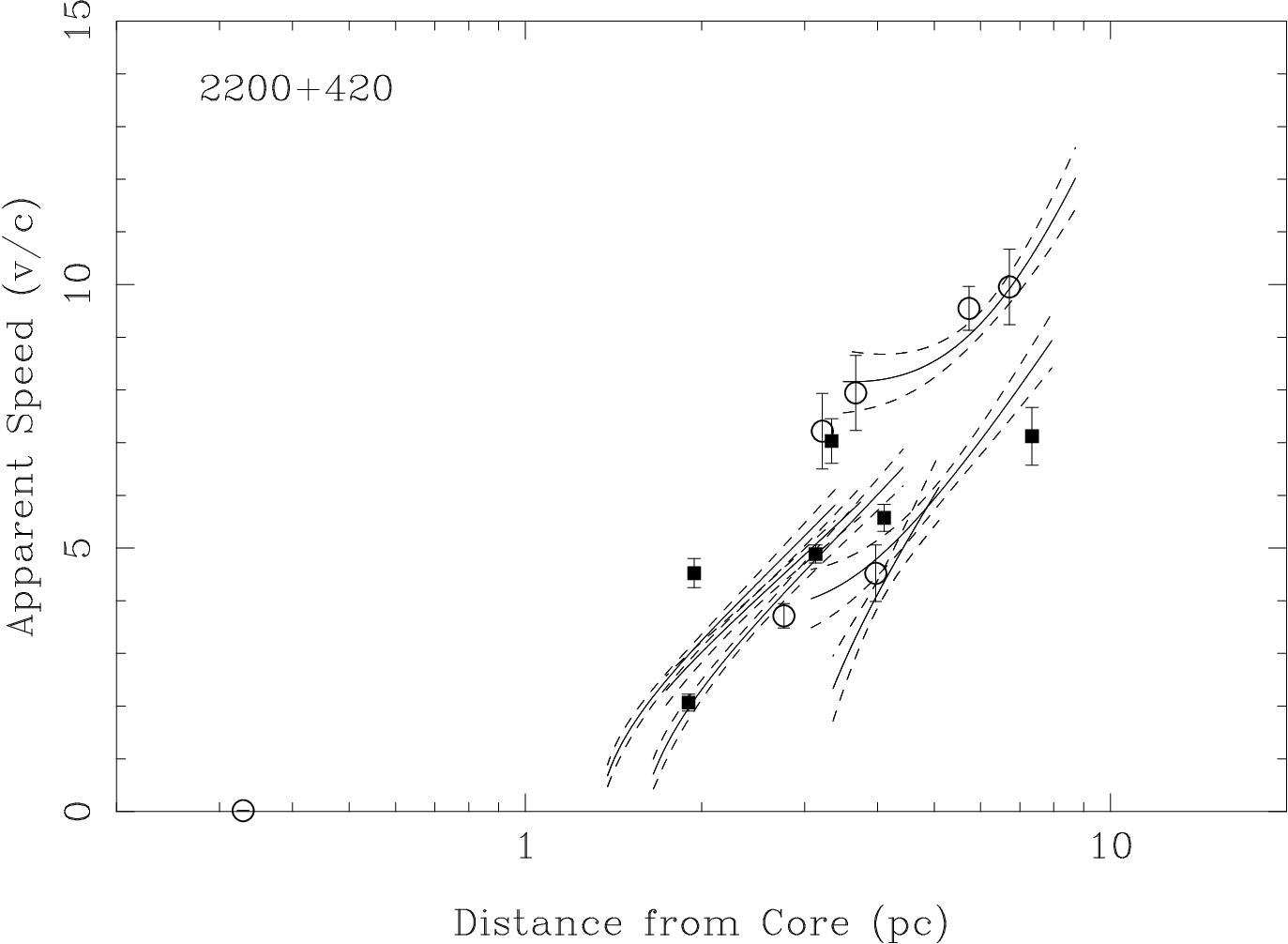}
\figcaption{
Figure continued from previous page.
}
\end{figure*}

\begin{figure*}
\figurenum{\ref{f:speed_v_distance}}
\includegraphics[scale=0.4,angle=-90,totalheight=0.19\textheight]{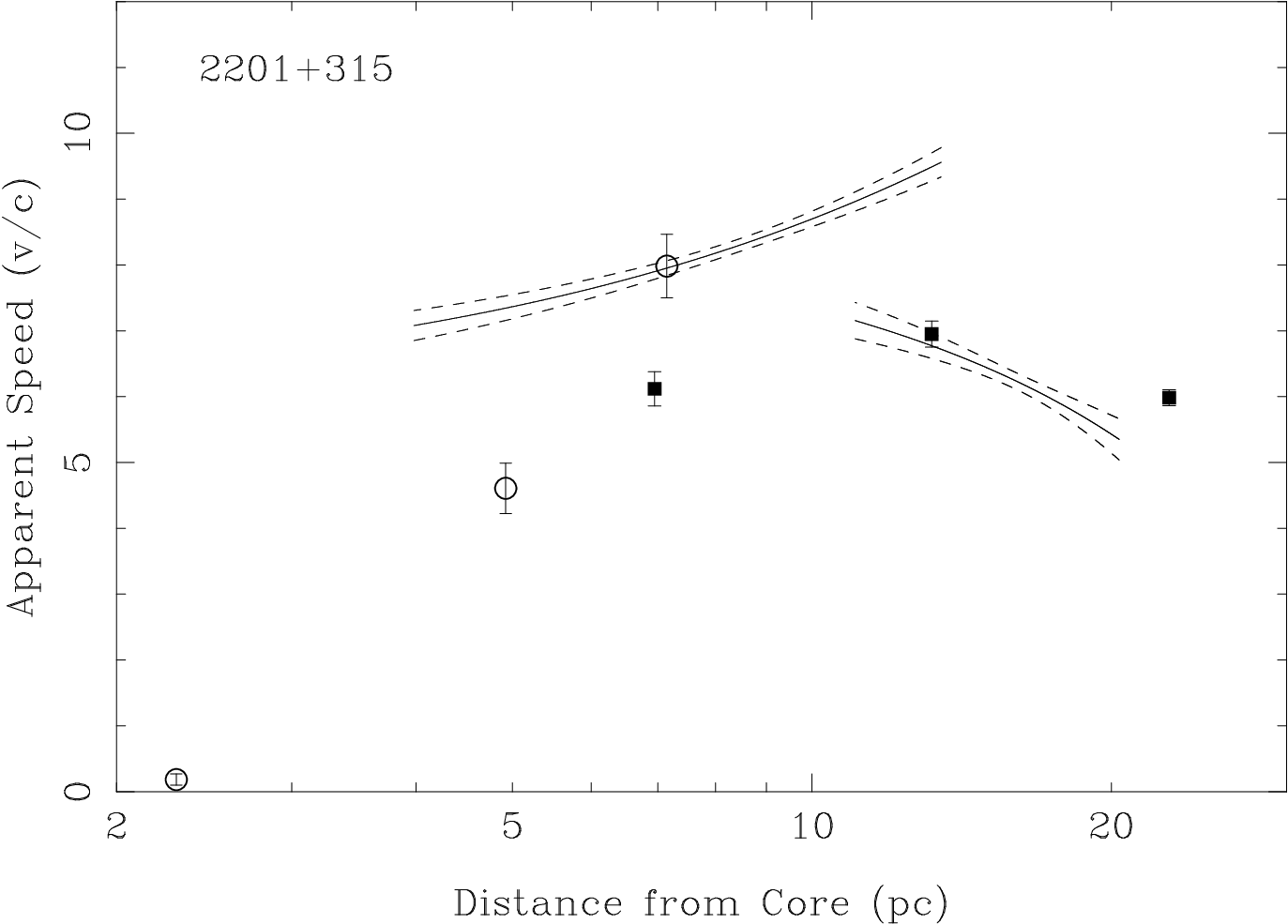}
\includegraphics[scale=0.4,angle=-90,totalheight=0.19\textheight]{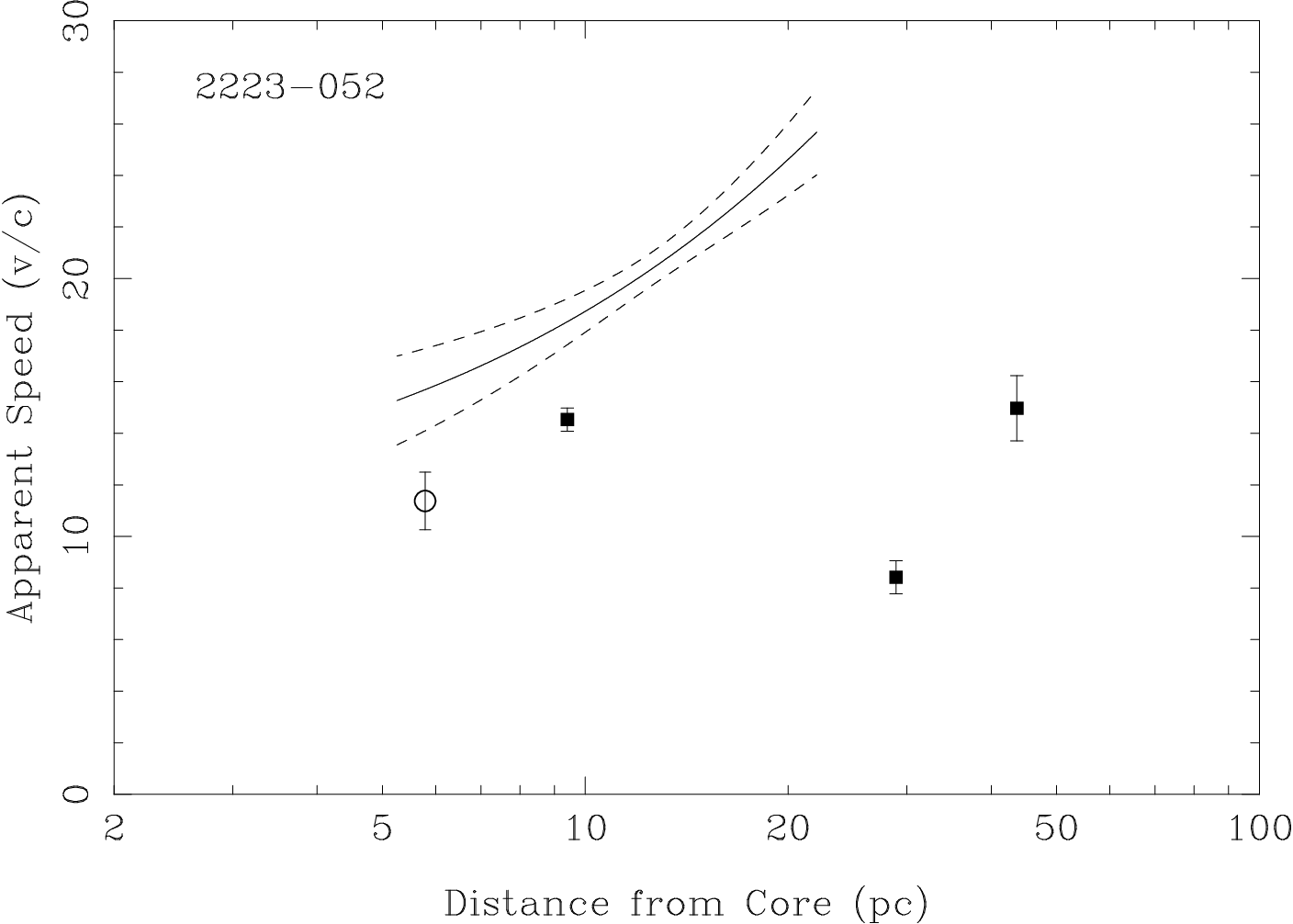}
\includegraphics[scale=0.4,angle=-90,totalheight=0.19\textheight]{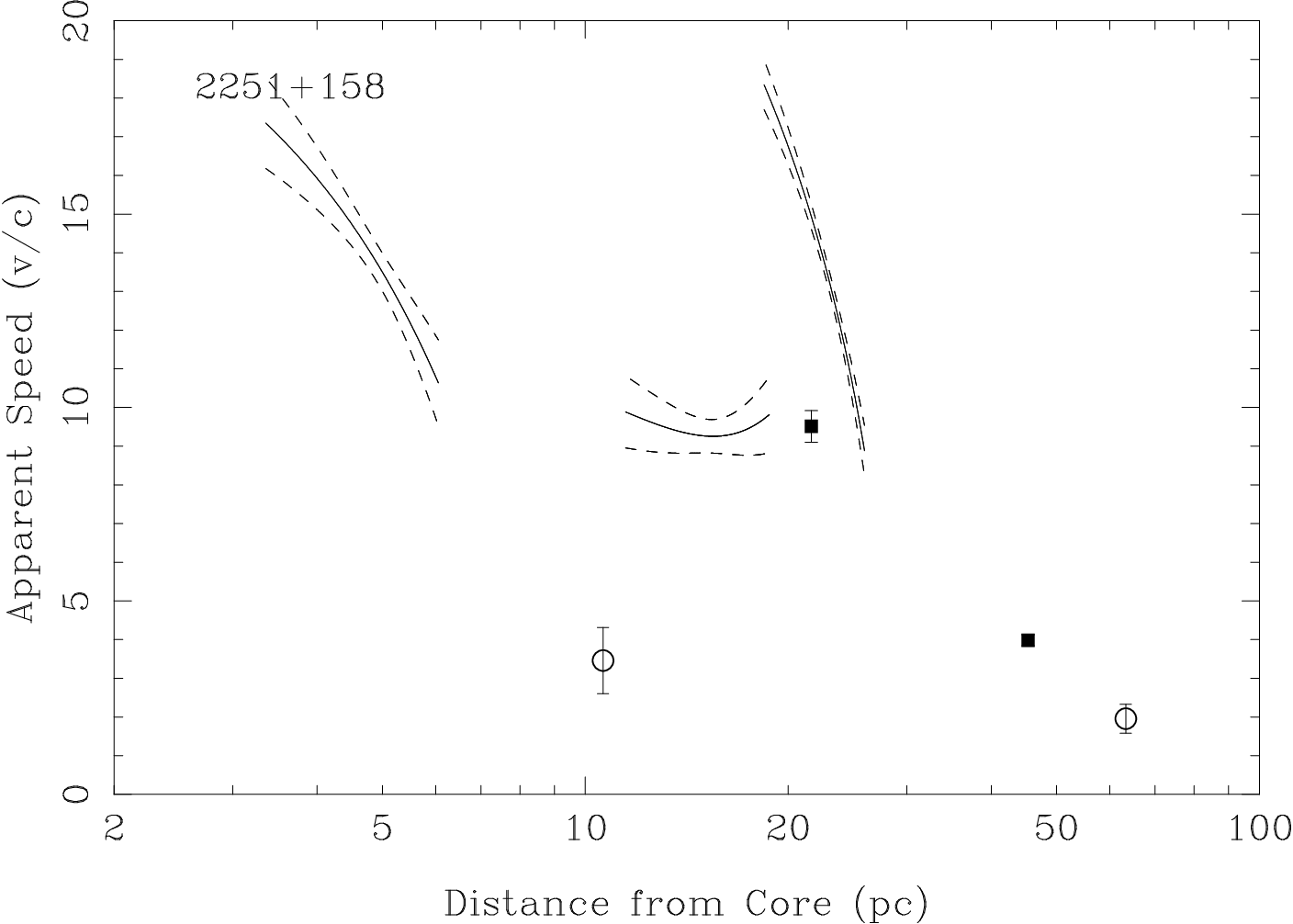}
\figcaption{
Figure continued from previous page.
}
\end{figure*}

As discussed in Paper X, jet
features in an individual source can have a large dispersion in apparent speeds.   
Figure \ref{f:speed_v_distance} shows that this dispersion is partly due to 
the observed acceleration and deceleration along the jet with distance; however, 
even at a fixed distance, there can be a large dispersion in apparent 
speeds in a given source, e.g. 1226$+$023 (3C\,273).  As shown in Paper X, viewing
angle differences between features in the same jet are expected
to cause some dispersion in apparent speed; however, we argue there that 
the conditions necessary to produce a large dispersion due to viewing angle alone
should also cause the jets to have a wide opening angle on the sky.  The same
conditions should also produce large perpendicular accelerations when jet
features change their trajectory, and we already conclude above that jet
features primarily change their apparent speed due to changes in their
Lorentz factor, not due to changes in their viewing angle. Hence we 
conclude that even at a fixed projected distance in an individual 
source, jet features can show a range of apparent speed due, at least in part, 
to differences in their Lorentz factors.

\begin{figure}
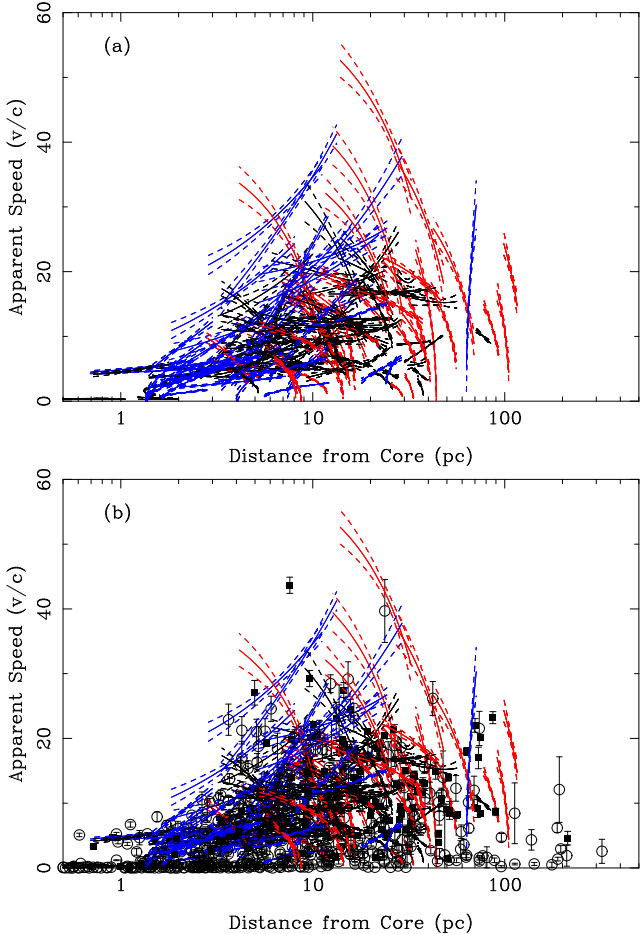

\includegraphics[scale=0.4,angle=-90,totalheight=0.26\textheight]{fig12a.eps}\\
\includegraphics[scale=0.4,angle=-90,totalheight=0.26\textheight]{fig12b.eps}
\figcaption{\label{f:speed_v_distance_all}
Plots of apparent speed vs. projected linear distance for all sources with known redshifts.  
In panel (a), 159 features with significant acceleration are plotted as linear projections of their fitted 
motion with one-sigma uncertainties defined by the upper and lower dashed lines.  Features plotted as 
blue and red trajectories have parallel/perpendicular acceleration ratios $\geq 2.0$, indicating
they are more likely to be due to real changes in Lorentz factor.  The colors, blue and red, 
indicate positive and negative parallel acceleration respectively for these 88 features, and
the rest are plotted in black.  
For clarity, six features in NGC 1052 with projected linear distances $<$ 0.5 pc are not plotted in panel (a).  
Panel (b) adds points representing the motion of 653 features for all sources with known redshifts in our sample.  
Features studied in this paper without significant accelerations are plotted as solid squares at the projected 
location of the midpoint of their motion. The apparent proper motion of the other robust features from 
Paper X are plotted as open circles.  A total of 40 features with projected linear distances
$<$ 0.5 pc and small motions are not plotted in panel (b).
}
\end{figure}

Finally, we consider the speed profile plot for our sample as a
whole.  Figure \ref{f:speed_v_distance_all}(a) collects the speed profiles
for all jet features showing significant parallel and/or perpendicular
accelerations.  Those accelerations that are most likely to be caused
by changes in Lorentz factor, as indicated by a parallel/perpendicular
acceleration ratio $\geq 2.0$, are plotted in either blue or red for
positive and negative accelerations, respectively.  Panel (b) includes
all the other features studied in this paper without significant
acceleration as solid squares and the
remaining robust features from Paper X as open circles.  The apparent
speed distribution has the largest values in the range $40-50$c at a 
projected linear distance of $\sim10$ pc.

\subsubsection{Lorentz Factors, Shocks, and The Jet Flow}
\label{s:shocks}

The Lorentz factors described above are for the moving jet features,
and they may not represent the Lorentz factor of the flow directly.  
Propagating shocks, proposed to explain both variability and VLBI jet features,
are expected to travel at a different speed from flow \citep[e.g.][]{HAA85,MG85},
and fast forward moving shocks can appear to move at two to three times
the speed of the flow \citep{CMA14,AHA14}. \citet{AGM01} show that a piston
driven shock, where there is an injection of new material into the flow with a higher 
than usual Lorentz factor, will produce a leading moving feature
that is nearly at the speed of the higher Lorentz factor; however, it will
also generate a number of slower, trailing shocks which appear to accelerate 
with the underlying background flow.  The picture is further complicated 
if one considers three dimensional effects of ejections of new 
moving features at multiple position angles \citep{AMG03}.  
Therefore a range of shock strengths and types, along with changes in
the injection speed or direction of new material, could plausibly account 
for the dispersion we see in jet feature speeds at a given distance in a jet.  

A traveling shock can change its speed if the underlying
conditions in the jet change, it dissipates its energy, or 
it interacts with other propagating disturbances in the flow.  To
produce the broad trends we see for accelerating features increasing
their speed at small distances and decreasing their speed at large
distances, a wide range of shock types and strengths across many
features and sources would need to change their speed in a similar
fashion over time and distance in the jet.
Given that shock speeds in the flow are added relativistically to 
the flow speed to determine the apparent speed of a moving 
feature, a natural common thread for a diversity of shock types
and strengths is the flow speed itself, as illustrated by the
trailing shocks in the numerical simulation of \citet{AGM01}.  We suggest that the 
acceleration of the flow itself is responsible for 
the broad trends we observe, and that the jet flow acceleration
region extends out to de-projected distances of order $10^2$ pc,
beyond which the flow begins to decelerate or remains constant.  

\subsection{Non-radial Motions and Jet Collimation}

Non-radial motions, or motions that do not extrapolate back to the
jet origin, comprise approximately half of our sample, indicating 
that jet features often do not follow ballistic trajectories. 
Observed perpendicular accelerations are almost always in the correct
direction to have caused the non-radial motion in that feature; 
however, we note that some jet features have non-radial motions without
significant perpendicular acceleration.  These features without
detectable perpendicular acceleration, or the few with perpendicular 
acceleration in the wrong direction, may have experienced earlier
periods of bending that changed their direction of motion prior 
to our observations.

Earlier work has shown that non-radial jet features typically have
their direction of motion rotated to better align with downstream emission 
\citep[][and Paper VII]{KL04},
indicating that the jet is following pre-established channels and jet features move
around the bends \citep{KL04}. In \S{3.2} we investigate the relationship
between non-radial motion and {\em upstream} emission, specifically the mean 
inner jet position angle.  We find that jet features that start out 
with a structural position offset from the mean inner jet direction tend to
experience kinematics which will bring them into better alignment.
This tendency of the motions 
to better align with the mean inner jet direction is strongest for jet features 
at small projected distances, $\lesssim 10$ pc from the core region, suggesting
that active collimation of the flow continues out to similar distances
as the acceleration discussed in the previous section, i.e. $\sim 10^2$ 
parsecs de-projected.  A connection between acceleration and collimation 
of the flow is expected in magnetic acceleration models which can 
continue to operate at large distances from the jet origin \citep[e.g.][]{VK04,KBV07,Ly09}.
Helical magnetic fields required by these models are suggested by a 
growing body of observational evidence from Faraday rotation \citep[e.g.][]{A02,
G04,A10,HLA12,ZSC13,G14}.

In a pure conically expanding flow, jet features would be free to 
move ballistically, so our result suggests that the jet width, $W_j$, is 
expanding less rapidly than $W_j \propto r$, unless jet features at large initial 
position angles have helical streaming motions that only make them appear to 
return toward the jet axis \citep[e.g.][]{HWG05}.  Only partial helical trajectories for individual
features have been observed to date \citep[e.g][]{SWV06,MAG14}, and the wavelength
of any such motion would have to be tuned to match the length scales observed here.  
Collimation of the jet itself is consistent with 
structural studies of M87 which show an edge-brightened jet \citep{J99,KLH07} with
an opening angle that narrows with distance from the core \citep{J99}. 
\citet{AN12} and \citet{HKD13} find $W_j \propto r^{0.6}$ for M87, and  
\citet{NHG14} find a more rapid collimation
profile in 3C\,84 of $W_j \propto r^{0.25}$.  We note that \citet{AN12} report that the 
jet of M87 maintains parabolic streamlines out to the location of HST-1 at 
a de-projected location of $\sim10^2$ parsecs where the maximum apparent speeds 
are observed, as discussed above.

\section{Summary and Conclusions}

We have analyzed acceleration and other kinematic data for 329
jet features in 95 sources. Our main findings are as follows:

(i) Accelerations and non-ballistic trajectories 
are common in the motions of jet features.  
Significant parallel accelerations, indicating a change in speed,
occur in 37\% of the jet features in our sample, and nearly
a quarter (23\%) have significant perpendicular accelerations indicating
a change in direction.  Nearly half (47\%) of our sample show
significant non-radial motion that does not extrapolate back to 
the jet origin.

(ii) In general parallel accelerations are distinctly larger 
than perpendicular accelerations.  The typical magnitudes
of parallel accelerations are nearly a factor of two times
the perpendicular accelerations, indicating that, on average,
changes to the Lorentz factors of jet features dominate the
observed parallel accelerations.

(iii) Parallel accelerations tend to be positive near the 
jet base, at short projected distance, and negative at 
longer distances.  The transition between speeding up
and slowing down seems to occur at roughly $\sim10-20$ pc
projected distance, with significant scatter.  Plots
of apparent speed versus distance confirm this broad 
relationship both for the entire sample and in individual 
sources, but again with significant scatter.  It is not
uncommon for the range of apparent speeds at a fixed distance 
in an individual jet to span a factor of two to three, indicating 
that a diversity of shock strengths and types may play a
role in determining the apparent motions of different features
in the same jet.  We argue that while different types of
shocks may explain the diversity of speeds at a given
distance, they cannot easily explain the overall trends
of individual features speeding up and slowing down across 
time and distance in the jet. We suggest that the common
thread that produces this trend is an overall acceleration 
of the jet flow out to de-projected distances of order
$10^2$ parsecs, beyond which the flow begins to 
decelerate or remains constant in speed.

(iv) For a typical Doppler beaming factor of $\delta\sim10$, we 
estimate the implied Lorentz factor changes of accelerating 
jet features fall in the range $\dot{\Gamma}/\Gamma \simeq 10^{-3}-10^{-2}$
per year in the host galaxy rest frame. While rates of change of this
order may appear small, they act over very long periods of time. 
A typical decade-long observation may correspond to centuries 
in the host galaxy due to the extreme time compression created by 
motion near our line of sight. As our plotted speed profiles
show, these intrinsic rates of change are sufficient to change 
apparent speeds by factors of two to three in many cases, indicating total 
Lorentz factor changes of a similar magnitude.

(v) The non-radial motion and perpendicular accelerations
of jet features that start out with large offsets from the
mean inner jet direction are typically in the correct 
direction to better align those features with the inner
jet direction.  This connection between directional changes and
upstream emission is strongest for jet features at 
small linear distances, $\lesssim 10$ pc projected, 
indicating the jet is still becoming collimated on 
length scales similar to the observed acceleration. 
However, we note that motion along helical stream
lines for individual jet features could reproduce this
result while still allowing the jet envelope to expand conically,
although conical expansion was not observed in structural studies of 
the nearby jets M87 or 3C\,84 \citep[e.g.][]{AN12,NHG14}.

\acknowledgements
Special thanks to the other members of the MOJAVE team for 
contributions to this project.
The MOJAVE project was supported under NASA-{\it Fermi} grants
NNX08AV67G and NNX12A087G.
DCH was supported by NSF grant AST-0707693. 
YYK was supported by the Russian Foundation for Basic Research 
(project 13-02-12103), 
Research Program OFN-17 of the Division of Physics, Russian Academy of Sciences,
and the Dynasty Foundation. ABP
was supported by the ``Non-stationary processes in the Universe''
Program of the Presidium of the Russian Academy of Sciences. 
TS was partly supported by the Academy of Finland project 274477.
ER was partially supported by the Spanish MINECO projects AYA2009-13036-C02-02 
and AYA2012-38491-C02-01 and by the Generalitat Valenciana project PROMETEO/2009/104, as well
as by the COST MP0905 action `Black Holes in a Violent Universe'.  
This work
made use of the Swinburne University of Technology software correlator
\citep{D11}, developed as part of the Australian Major
National Research Facilities Programme and operated under licence.
This research has made use of NASA's Astrophysics Data System.



%
%

%
%

\clearpage

\LongTables
\setlength{\tabcolsep}{2pt}
\begin{deluxetable*}{llrrrrrrrrrrrr} 
\tablecolumns{14} 
\tabletypesize{\scriptsize} 
\tablewidth{0pt}  
\tablecaption{\label{t:comps} Jet Features for Acceleration and Non-Radial Motion Analysis}  
\tablehead{
\colhead{} &   \colhead {} & \colhead{} & \colhead{$\langle R\rangle$} &
\colhead{$\langle\vartheta\rangle$} & \colhead {$d_{proj}$} &   \colhead {$\mu$} &  \colhead{$\beta_{obs}$} &
\colhead{$\phi$} & \colhead{$|\langle\vartheta\rangle-\phi|$} &
\colhead{$\dot{\mu}_\parallel$} & \colhead {$\dot{\mu}_\perp$} &   \colhead{$\dot{\eta}_\parallel$} & \colhead{$\dot{\eta}_\perp$} \\
\colhead{Source} &   \colhead {I.D.} & \colhead{N} & \colhead{(mas)} &
\colhead{(deg.)} & \colhead {(pc)} &   \colhead {($\mu$as yr$^{-1}$)} & & \colhead{(deg.)} & \colhead{(deg.)} &
\colhead{($\mu$as yr$^{-2}$)} & \colhead {($\mu$as yr$^{-2}$)} &   \colhead {(yr$^{-1}$)} & \colhead{(yr$^{-1}$)} \\
\colhead{(1)} & \colhead{(2)} & \colhead{(3)} & \colhead{(4)} &  \colhead{(5)} &
\colhead{(6)} & \colhead{(7)} & \colhead{(8)} & \colhead{(9)} &  \colhead{(10)} &
\colhead{(11)} & \colhead{(12)} & \colhead{(13)} & \colhead{(14)} 
}
\startdata 
0003-066 & $4$ & $23$ & $6.63$ & $-74.6$ & $32.34$ & $80\pm6$ & $1.7\pm0.1$ &  $-90.3\pm2.8$ & $15.6\pm2.8$ & $-24\pm3$ & $10\pm2$ & $-0.41\pm0.06$ & $0.17\pm0.03$ \\
0010+405 & $1$ & $10$ & $8.18$ & $-31.3$ & $32.26$ & $428\pm40$ & $6.9\pm0.6$ &  $-19.2\pm4.1$ & $12.1\pm4.1$ & $-43\pm69$ & $11\pm53$ & $-0.13\pm0.20$ & $0.03\pm0.16$ \\
0016+731 & $1$ & $15$ & $1.19$ & $140.3$ & $10.15$ & $106\pm4$ & $8.2\pm0.3$ &  $163.2\pm2.2$ & $22.8\pm2.4$ & $9\pm2$ & $10\pm2$ & $0.23\pm0.05$ & $0.27\pm0.05$ \\
0059+581 & $2$ & $21$ & $1.70$ & $-116.3$ & $11.69$ & $168\pm14$ & $6.2\pm0.5$ &  $-102.0\pm3.3$ & $14.3\pm3.5$ & $-16\pm8$ & $8\pm5$ & $-0.16\pm0.08$ & $0.08\pm0.05$ \\
         & $3$ & $14$ & $0.92$ & $-124.7$ & $6.34$ & $185\pm13$ & $6.8\pm0.5$ &  $-114.9\pm3.2$ & $9.8\pm3.4$ & $-83\pm15$ & $-16\pm12$ & $-0.73\pm0.14$ & $-0.14\pm0.11$ \\
         & $4$ & $14$ & $0.67$ & $-157.4$ & $4.61$ & $154\pm11$ & $5.7\pm0.4$ &  $-125.1\pm4.0$ & $32.3\pm4.2$ & $-2\pm14$ & $12\pm14$ & $-0.02\pm0.15$ & $0.13\pm0.15$ \\
         & $5$ & $13$ & $0.53$ & $-163.4$ & $3.65$ & $239\pm8$ & $8.9\pm0.3$ &  $-120.8\pm2.9$ & $42.5\pm3.0$ & $103\pm13$ & $61\pm19$ & $0.71\pm0.09$ & $0.42\pm0.13$ \\
0202+149 & $5$ & $11$ & $3.11$ & $-66.2$ & $16.72$ & $568\pm53$ & $14.0\pm1.3$ &  $-56.8\pm4.3$ & $9.4\pm4.6$ & $-44\pm45$ & $21\pm36$ & $-0.11\pm0.11$ & $0.05\pm0.09$ \\
\enddata
\tablecomments{
Columns are as follows: 
(1) Source name in B1950 coordinates; 
(2) Component ID;
(3) Number of epochs;
(4) Mean radial separation from core in milli-arcseconds (averaged over all epochs);
(5) Mean structural position angle in degrees;
(6) Mean projected radial distance in parsecs;
(7) Angular proper motion in micro-arcseconds per year;
(8) Apparent speed in units of the speed of light;
(9) Proper motion position angle in degrees;
(10) Absolute difference between mean structural position angle and proper motion position angle in degrees;
(11) Angular acceleration parallel to the proper motion position angle in micro-arcseconds per year per year;
(12) Angular acceleration perpendicular to the proper motion position angle in micro-arcseconds per year per year;
(13) Relative parallel acceleration as defined in \S{3};
(14) Relative perpendicular acceleration as defined in \S{3};   
}
\end{deluxetable*}

\clearpage

\end{document}